\documentclass[aip,pop,preprint,a4paper]{revtex4-1}

\usepackage{amsmath}
\usepackage{amssymb}
\usepackage{graphicx}% Include figure files
\usepackage{dcolumn}% Align table columns on decimal point
\usepackage{bm}% bold math

\draft % marks overfull lines with a black rule on the right

\begin{document}

%\preprint{}

\title{Automation of The Guiding Center Expansion} %Title of paper



\author{J. W. Burby}
 \affiliation{Princeton Plasma Physics Laboratory, Princeton, New Jersey 08543, USA}
\author{J. Squire}
 \affiliation{Princeton Plasma Physics Laboratory, Princeton, New Jersey 08543, USA}
\author{H. Qin}
 \affiliation{Princeton Plasma Physics Laboratory, Princeton, New Jersey 08543, USA}
 \affiliation{Dept. of Modern Physics, University of Science and Technology of China, Hefei, Anhui 230026, China}
%\thanks{}


\date{\today}

\begin{abstract}
We report on the use of the recently-developed Mathematica package \emph{VEST} (Vector Einstein Summation Tools) to automatically derive the guiding center transformation. Our Mathematica code employs a recursive procedure to derive the transformation order-by-order. This procedure has several novel features. (1) It is designed to allow the user to easily explore the guiding center transformation's numerous non-unique forms or representations. (2) The procedure proceeds entirely in cartesian position and velocity coordinates, thereby producing manifestly gyrogauge invariant results; the commonly-used perpendicular unit vector fields $e_1,e_2$ are never even introduced. (3) It is easy to apply in the derivation of higher-order contributions to the guiding center transformation without fear of human error. Our code therefore stands as a useful tool for exploring subtle issues related to the physics of toroidal momentum conservation in tokamaks. 
\end{abstract}


\maketitle %\maketitle must follow title, authors, abstract and \pacs

\section{Introduction}

  The guiding center asymptotic expansion is both beautiful and revolting. Its beauty stems from its simple physical underpinning; a strongly magnetized charged particle gyrates around magnetic field lines much more rapidly than it drifts along or across them. This simplicity allows the approximation to be applied in a greater variety of settings than perhaps any other approximation scheme used in magnetized plasma physics. And in spite of the approximation's broad appicability, which might be expected to dilute its power, it affords significant practical benefits. Perhaps most notably, it enables gyrokinetic codes, such as those discussed in Refs. \onlinecite{jenko} and \onlinecite{xgc1}, to work on the drift, rather than gyroperiod, time scale.

The approximation begins to reveal its ugly side, however, when one endeavors to derive successively higher-order contributions to the expansion\cite{northrop,BrTr}. Aside from the usual proliferation of terms common amongst higher-order perturbation expansions, the obstacles one encounters include vector identities involving spatially varying unit vectors such as $b=\mathbf{B}/|B|$ and subtle issues related to \emph{gyrogauge invariance}\cite{GGI}. Moreover, attempts to taylor the expansion to respect the Hamiltonian structure of the Lorentz force law encounter the so-called \emph{order-mixing}\cite{BrTr,Brizard_thesis} issue, whereby different components of the coordinate transformation one seeks appear at different orders in the transformed Lagrangian, thus complicating the procedure used to find them.

These abhorrent features can be frightening to the uninitiated. As a result, only a dedicated minority have ever attempted delving into the calculation beyond the derivation of drifts proportional to first derivatives of the magnetic field. The reluctant majority, up until fairly recently\cite{krommes12}, could have justified their stance by proclaiming the higher-order corrections to be practically unimportant, and therefore irrelevant. Recent advances, however, are making it more and more clear that at least corrections proportional to second derivatives of the magnetic field are important for resolving the physics of toroidal momentum conservation in tokamaks\cite{PaCa10}. For this reason, certain largely unexplored aspects of these higher-order corrections now appear intriguing to study. In particular, the various \emph{representations} of the guiding center expansion should be explored further. 

A representation of the guiding center expansion consists of a prescription for making all of the apparently arbitrary choices one must make in the process of deriving the expansion. Examples of different representations can be found In Ref. \onlinecite{BrTr}, where two representations are presented, or in Littlejohn's work in Refs. \onlinecite{littlejohn81} and \onlinecite{littlejohn83}. There is nothing unphysical about these different representations - they merely arise from the fact that equations of motion which are independent of gyrophase will remain so upon an arbitrary coordinate transformation that commutes with the gyrosymmetry operation (see appendix \ref{appA}). Nevertheless, different representations lead to guiding center equations of motion with different numbers of terms. Thus, one could imagine optimizing the number of terms in the equations of motion over the space of representations. It is also possible that different representations  have different times of validity. After all, Kruskal's method\cite{kruskal62}, which provides the mathematical basis for the guiding center expansion, can only guarantee equations of motion valid for times of order $1/\epsilon$, where $\epsilon$ is the ordering parameter $\rho/L$\cite{omohundro}.

In order to enable the study of these issues, a process which would surely involve deriving the guiding center expansion in many different representations, we have developed, implemented, and verified an algorithm to automate the guiding center calculation using the newly-developed Mathematica package \emph{VEST} (Vector Einstein Summation Tools)\cite{SqBuQi13}. In particular, we have slashed the time required to derive the expansion, and all but eliminated the possible taint of human-made algebra errors in the derivation of higher-order contributions to the guiding center expansion.

 While other authors have presented algorithmic procedures for deriving the guiding center expansion in the past \cite{kruskal58,larsson,BrTr}, the algorithm we present here is novel due to the combination of the following.

\noindent 1) The algorithm \emph{has actually been implemented on a computer} and used to derive the guiding center expansion in two different representations.

\noindent 2) Complicated, multi-term, vector identities are accounted for using the clever simplification capabilities of \emph{VEST}.

\noindent 3) Issues related to gyrogauge invariance are completely avoided by working in cartesian position and velocity coordinates. In particular, the only unit vector that plays a role is the physical $b=\mathbf{B}/|B|$.

\noindent 4) Gyroaverages and Fourier expansions in gyrophase are implemented in these coordinates using a coordinate-independent formulation of these operations.

\noindent 5) The approach manages to be manifestly Hamiltonian while addressing the order-mixing issue in a computationally attractive manner; for each $m>n$, the $n$'th-order contribution to the perturbative coordinate transformation is determined without knowledge of any of the details of the $m$'th order contribution.

\noindent 6) The manner in which we address the order-mixing issue obviates the high degree of freedom in the form of the transformed Lagrangian.

In what follows, we will describe our algorithm and report on the equations of motion generated in the two representations just alluded to. We will \emph{not} evaluate these new representations in terms of their simplicity or time-validity properties; a properly thorough study of these properties will appear in future work. We will begin with four sections describing what our algorithm is meant to do as well as our motivation for selecting an algorithm with the novel features just described. In section \ref{two}, we give a schematic overview of Hamiltonian Lie transform-based perturbation theory in order to remind the reader of the goal of the guiding center expansion. We then describe the motivation for selecting our algorithm via a description of three difficulties we faced while developing it, and how we overcame them. In particular, sections \ref{prime}, \ref{three}, and \ref{four} are devoted to discussing the difficulties presented by the order-mixing issue; the desire for manifestly gyrogauge invariant results; and the task of computing gyroaverages and gyroharmonics, respectively. With all of the motivations in place, we  present our algorithm in section \ref{five}. Finally, in section \ref{six}, we present the results of automatically performing the guiding center expansion with our algorithm in two previously unstudied representations.

\section{\label{two}A Schematic for Hamiltonian Lie Transform Perturbation Theory}
In this section, we will review the general structure and purpose of the guiding center expansion, and thereby indicate precisely what our algorithm is meant to do. We then discuss three key difficulties we faced while trying to develop the algorithm before actually presenting presenting it. The purpose of these first four sections is to provide a narrative explaining why the algorithm looks the way it does. Readers only interested in the algorithm itself can skip straight to section \ref{five}, but it may still be useful to skim these early sections in order to become familiar with our notation. 

We begin by recalling the coordinate-independent formulation of Hamiltonian dynamical systems\cite{FoM}. This formulation makes use of Cartan's exterior calculus of differential forms; a very brief overview of the latter is provided in Appendix \ref{appLie}. The phase space $M$ is assumed to be an even dimensional smooth manifold\cite{smooth_manifolds} equipped with a symplectic two-form $\omega$. The dynamical equations are then specified by a function $H:M\rightarrow\mathbb{R}$ known as the Hamiltonian function via Hamilton's equations
\begin{align}\label{hamilton}
\text{i}_{X_H}\omega=\mathbf{d}H,
\end{align}  
where $X_H$ is the vector field that specifies the time derivative of any particle's phase space location $c(t)\in M$, i.e. $c^\prime(t)=X_H(c(t))$.
In any local coordinate system $(z^i)$ on $M$, Hamilton's equations become
\begin{align}
\dot{z}^i\omega_{ij}=\frac{\partial H}{\partial z^j},
\end{align} 
where $\dot{z}^i$ are the components of the vector field $X_H=\dot{z}^i\frac{\partial}{\partial z^i}$ and $\omega_{ij}=\omega(\frac{\partial}{\partial z^i},\frac{\partial}{\partial z^j})$. 

In the guiding center problem, the phase space is the six-dimensional position-velocity space, $M=\mathbb{R}^3\times\mathbb{R}^3$, equipped with the symplectic form $\omega_\epsilon=-\mathbf{d}\vartheta_\epsilon$, where the one-form $\vartheta_\epsilon$ is given in terms of the the magnetic vector potential $\mathbf{A}$ and the guiding center ordering parameter $\epsilon$ by
\begin{align}\label{lorentz_one_form}
\vartheta_\epsilon=\mathbf{A}\cdot dx+\epsilon v\cdot dx.
\end{align}
The equations of motion are then specified by the Hamiltonian function $H=\frac{1}{2}\epsilon^2v\cdot v$. As can be readily verified, the vector field $X_H(\epsilon)$ in the natural cartesian coordinates on $M$ is given by
\begin{align}\label{lorentz}
\dot{v}(x,v)&=v\times\mathbf{B}(x)\\
\dot{x}(x,v)&=\epsilon v.\nonumber
\end{align}
Strictly speaking, the placement of the ordering parameter $\epsilon=\rho/L$ in $\vartheta_\epsilon$, and therefore its placement in the equations of motion, is only justified in appropriate dimensionless variables, as discussed in Ref. \onlinecite{northrop}. However, we can regard Eq.\,(\ref{lorentz}) as a dimensional equation if we think of $\epsilon$ as a formal ordering parameter and if we normalize $\mathbf{A}$ by a particle's charge-to-mass ratio so that $\mathbf{B}$ has the dimension of frequency.

When $\epsilon=0$, which corresponds to the asymptotic limit where a particle undergoes gyromotion with zero gyroradius and vanishingly slow drift, the equations of motion given in Eq.\,(\ref{lorentz}) are gyrosymmetric\cite{BuQi12,qinRINGS} (see section \ref{three} for the precise definition of gyrosymmetric tensors). Because the particle trajectories are periodic in this limit, Kruskal's general theory\cite{kruskal62} tells us that we can asymptotically deform, or rearrange, the phase space $M$ using a non-unique $\epsilon$-dependent near-identity transformation $T_\epsilon:M\rightarrow M$ such that the transformed $X_H(\epsilon)$, $X_H(\epsilon^\prime)\equiv T_{\epsilon*}X_H(\epsilon)$, is gyrosymmetric to \emph{all orders in} $\epsilon$. 

The goal of the guiding center theory, and therefore the algorithm we will present later, is to find such a $T_\epsilon$. Because performing this task requires a degree of ingenuity, a number of useful methods have been developed. Of particular relevance to the present work are those methods that employ Lie transforms. In these cases, one posits that the desired transformation from the old phase space to the new, deformed phase space can be expressed in the form\footnote{For some quick intuition regarding this ansatz, recall that the time-advance map associated to an arbitrary vector field $Y:M\rightarrow TM$ is given by $\exp(t Y)$. Thus, given $x\in M$, $T_\epsilon(x)$ is found by sequentially flowing along the vector fields $G_n(\epsilon)$ for $-1$ unit of time each, starting from $x$.}
\begin{align}\label{gen_trans}
T_\epsilon=...\circ\exp(-G_n(\epsilon))\circ...\circ\exp(-G_1(\epsilon)),
\end{align}
where, for each $n$, $G_n(\epsilon):M\rightarrow TM$ is a vector field that tends to zero as $\epsilon\rightarrow 0$, and does so more rapidly than does $G_m(\epsilon)$ with $m<n$.  The requirement that the transformed equations of motion be gyrosymmetric then reduces to a sequence of requirements on the $G_n(\epsilon)$. Thus, the Lie transform approach to finding $T_\epsilon$ reduces to finding a sequence of $G_n(\epsilon)$ that satisfy the latter requirements.

One can derive these requirements in one of two ways. The direct method, which recently made an appearance in Ref. \onlinecite{guillebon} (also see Ref. \onlinecite{KBM}), consists of formally computing the transformed $X_H(\epsilon)$, $X_H^\prime(\epsilon)=T_{\epsilon*}X_H(\epsilon)$, using Eq.\,(\ref{gen_trans}) and then demanding that the result be gyrosymmetric to all orders. The Hamiltonian method, due to Littlejohn\cite{littlejohn82}, consists of calculating the transformed $\mathbf{d}\vartheta_\epsilon$ and $H$, $T_{\epsilon*}\mathbf{d}\vartheta_\epsilon$ and $T_{\epsilon*}H$, and then demanding that each of be gyrosymmetric to all orders. These two methods are related by the general fact that if the symplectic form, $\omega$, and Hamiltonian, $H$, appearing in Hamilton's equations (\ref{hamilton}) admit a symmetry, then so does $X_H$. Each method also involves making a number of arbitrary decisions to completely determine the $G_n(\epsilon)$; different choices lead to different representations.  

In principle, either approach can be automated on a computer. Indeed, historically this has been one of the advertised ``features" of the Lie transform approach to perturbation problems in general. However, the Hamiltonian approach has the advantage of providing an attractive means for truncating the results of the expansion. Generally, given a near-identity transformation and a Hamiltonian system specified by a one-form and Hamiltonian, there are two ways to develop ``finite" approximations, or truncations to the transformed equations of motion. One approach is to directly truncate the transformed equations of motion at some order in the expansion parameter. The other approach is to truncate the transformed one-form and Hamiltonian and use the vector field specified by the ensuing Hamilton's equations to approximate the transformed equations of motion. Either approach may be used to generate arbitrarily accurate approximations to the transformed equations of motion. However, the second approach always produces an approximation to the transformed equations of motion that is rigorously Hamiltonian. Because the original dynamical system is Hamiltonian, the second ``Hamiltonian" truncation scheme is theoretically preferable. Thus, an advantage offered by the Hamiltonian approach to deriving the guiding center transformation is as follows. Because the Hamiltonian approach directly tracks the one-form and Hamiltonian through the near-identity transformation (\ref{gen_trans}), the Hamiltonian truncation scheme is always immediately available; it is not necessary to compute the transformed one-form and Hamiltonian \emph{after} finding the $G_n(\epsilon)$ as it would be if the $G_n(\epsilon)$ were calculated using the direct method. It is for this reason that we pursue the Hamiltonian approach in the present work.

\section{\label{prime}Difficulty 1: order-mixing}
While developing the algorithm for automating the Hamiltonian Lie transform approach to finding $T_\epsilon$, we encountered three key difficulties. In this section and the two that follow we will describe each in turn, as well as the manner in which we overcame each difficulty.
 
The first issue is rooted in the special form of $\vartheta_\epsilon$ given above. If one specifies the $\epsilon$-dependence of the $G_n(\epsilon)$ according to $G_n(\epsilon)=\epsilon^n g_n$, then the one-form $\vartheta_\epsilon$ on the deformed phase space is given by 
\begin{align}
\vartheta_\epsilon^\prime= &\mathbf{A}\cdot dx +\epsilon v\cdot dx +\epsilon L_{g_1}( \mathbf{A}\cdot dx )+O(\epsilon^2),
\end{align}
where $L_{g_1}$ denotes the Lie derivative with respect to the vector field $g_1$ (see appendix \ref{appLie}). In order to find one of the transformations guaranteed by Kruskal's theory, the combination $v\cdot dx + L_{g_1}( \mathbf{A}\cdot dx )$ must be gyrosymmetric, modulo closed one-forms\footnote{Demanding the transformed $\mathbf{d}\vartheta_\epsilon$ to be gyrosymmetric is equivalent to demanding that the transformed $\vartheta_\epsilon$ be equal to the sum of a gyrosymmetric one-form and an arbitrary closed one-form. This follows from the fact that $\mathbf{d}\vartheta_\epsilon^\prime$ is left unchanged upon replacing $\vartheta_\epsilon^\prime$ with $\vartheta_\epsilon^\prime+\alpha$ for an arbitrary closed one-form $\alpha$, i.e. $\mathbf{d}\alpha=0$}. If we write $g_1=g_1^x\cdot\frac{\partial}{\partial x}+g_1^v\cdot\frac{\partial}{\partial v}$, this condition can be satisfied by choosing $g_1^x=\frac{1}{|B|}v\times b+\alpha b$, $g_1^v=Y$, where $\alpha$ and $Y$ are arbitrary. However, as would become clear upon analyzing higher-order contributions to the transformed one-form, there are in fact constraints on $\alpha$ and $Y$, meaning at least part of the Freedom in specifying $g_1$ suggested by the first order change in the one-form is only apparent. This is a special case of the more general order-mixing issue; the constraints on a given $g_n$ can only be deduced by considering multiple orders $\vartheta_\epsilon^\prime$, in particular orders whose form is effected by $g_m$ for $m>n$. 
   
While order-mixing does not prevent the success of the Hamiltonian method (see Ref. \onlinecite{BrTr} for one way of coping with it), from a computational point of view, it is bothersome. It obfuscates the extent to which the choices one needs to make to find the various $g_n$ are coupled across $n$-values. If the coupling were severe enough, then any algorithm one might construct to automate these choices could be very complicated. 

In order to overcome this difficulty, we designed our algorithm to satisfy 
\\ \\
\noindent \emph{Resolution of Difficulty 1:}  
The rule for determining $G_n(\epsilon)$ does not rely on any specific knowledge of any component of $G_m(\epsilon)$ whenever $m>n$.
\\ \\
In section \ref{five}, the precise manner in which the algorithm accomplishes this will become clear.

\section{\label{three}Difficulty 2: Manifest Gyrogauge Invariance}
\begin{figure}
\includegraphics{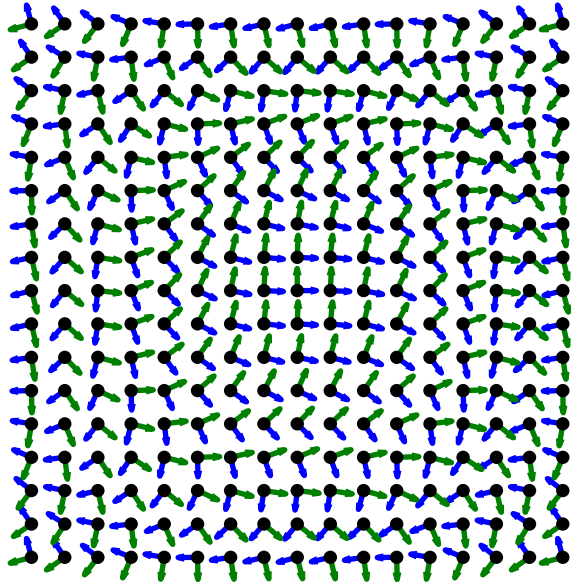}
\caption{\label{perp}A typical arrangement of the perpendicular unit vectors $e_1,e_2$ for a uniform magnetic field that points out of the page. The two sets of arrows represent $e_1$ and $e_2$. While in this case, $e_1$ and $e_2$ are not \emph{required} to vary in space, for a more general sort of magnetic field, they would be. Reprinted from Phys. Plasmas 19, 052106 (2012). Copyright 2012 American Institute of Physics.}
\end{figure}

In order to understand the second difficulty we faced in developing a good algorithm for automating the guiding center calculation, one needs to understand the usual definition of a gyrosymmetric tensor. This definition refers to a special type of coordinate system on $M$, any instance of which we will call a \emph{fibered coordinate system}. A fibered coordinate system on $M$ consists of an open subset $U\subset M$, $5$ smooth functions $\xi^i:U \rightarrow \mathbb{R}$, $i=1,...,5$, and one additional function $\theta:U\rightarrow \mathbb{R}~\text{mod}~2\pi$ satisfying:
\\ \\
F1\footnote{F stands for ``fibered''.}: The six functions $\xi_i$, $i=1,...,5$, and $\theta$ define a valid coordinate system on $U$
\\ \\
F2: Holding the $\xi_i$ fixed, $\theta$ parameterizes, in a left-handed sense relative to $b$, the zero'th order ($\epsilon=0$) solutions to Eq.\,(\ref{lorentz}), which are called \emph{loops} by Kruskal. 
\\ \\
The standard example of a family of fibered coordinate systems used in guiding center theory is constructed as follows. First find a smooth unit vector field $e_1$ perpendicular to the magnetic field, $e_1\cdot b=0$. $e_1(x)$ and $e_2(x)=(b\times e_1)(x)$ span the plane perpendicular to $b(x)$ for each $x$ in the domain of definition, $D\subset\mathbb{R}^3$, of $e_1$, as depicted in Figure \ref{perp}. As shown in Ref. \onlinecite{BuQi12}, $D$ cannot always be taken to be the entire $3$-dimensional domain particles move through. A fibered coordinate system can then be defined on the open subset of phase space $U\equiv \left\{(x,v)\in M|x\in D~\text{and}~b(x)\times v\neq 0\right\}$. Labeling the $\xi_i$ according to $(\xi_1,\xi_2,\xi_3)=x,~\xi_4=v_\perp,~\xi_5=v_\parallel$, these functions are defined by the relation
\begin{align}
x=&x(x,v)\\
v=&v_\parallel(x,v)b(x)\nonumber\\
+&v_\perp(x,v)\left(\cos(\theta(x,v))e_1(x)-\sin(\theta(x,v))e_2(x)\right).\nonumber
\end{align}

A gyrosymmetric tensor can then be defined in terms of fibered coordinate systems as follows. A tensor is gyrosymmetric if its components in an arbitrary fibered coordinate system do not depend on $\theta$. Note that one doesn't have to look at a tensor in every fibered coordinate system to check this property; it is enough to check in a collection of fibered coordinate systems that cover $M$.

This standard definition of gyrosymmetric tensors motivates the standard approach to deriving the constraints on the $G_n(\epsilon)$. One first writes out the components of the transformed $\mathbf{d}\vartheta_\epsilon$ and $H$ in a family of fibered coordinate systems on $M$ that cover $M$. Then one chooses the local representatives of $G_n(\epsilon)$ to eliminate the $\theta$-dependence in these components in each coordinate system in the covering. For consistency\cite{BuQi12}, one also must demand that the local definitions of $G_n(\epsilon)$ agree when changing from one fibered coordinate system in the covering to another. This last consistency condition is one statement of the principle of gyrogauge invariance. Satisfying this consistency condition is equivalent to demanding that the local representatives of $G_n(\epsilon)$ have the gyrogauge invariant form first identified by Littlejohn in Ref. \onlinecite{GGI}.

There is nothing conceptually wrong with this approach to finding the $G_n(\epsilon)$, and it can be made to work. However, there is a very practical problem with proceeding in precisely this manner on a computer. In order to verify that a given expression for $G_n(\epsilon)$ in a fibered coordinate system satisfies the principle of gyrogauge invariance, it is often necessary to account for non-trivial vector identities involving the perpendicular unit vectors $e_1$ and $b$. For instance, the identity
\begin{align}\label{N}
\nabla\times((\nabla e_1)\cdot e_2)=&\frac{1}{2}b\bigg(\text{Tr}(\nabla b\cdot \nabla b)-(\nabla\cdot b)^2\bigg)\\
&+(\nabla\cdot b)b\cdot\nabla b-b\cdot\nabla b\cdot \nabla b,\nonumber
\end{align}
and even more complicated identities generated by taking derivatives of Eq.\,(\ref{N}) must be recognized. Presently, there is no general method that would allow one to do this on a computer in all cases one might encounter. Thus, one cannot guarantee that the $G_n(\epsilon)$ produced by a computer following the above procedure will \emph{manifestly} exhibit gyrogauge invariance, i.e. it will not be obvious that $G_n(\epsilon)$ is  gyrogauge invariant, even if it actually is.

In order to avoid this issue, we have chosen to avoid using fibered coordinate systems altogether. 
\\ \\
\emph{Resolution of Difficulty 2:}
All tensors are expressed and manipulated in cartesian position and velocity coordinates.
\\ \\
By proceeding in this manner, the results generated by our algorithm ($G_n(\epsilon)$, for example) will be expressed entirely in terms of $v$, $|B|$, $b$, and derivatives thereof, thereby making our algorithm manifestly gyrogauge invariant; perpendicular unit vectors $e_1,e_2$ and the gyrophase coordinate $\theta$ are never even introduced.

\section{\label{four}Difficulty 3: Fourier Analysis Without Introducing Additional Unit Vectors}
The third issue we wish to discuss arises as a result of our resolution of the difficulty discussed in the previous section. Because part of the motivation for choosing to work in Cartesian coordinates was to avoid introducing additional unit vectors, it would be a step backward if we had to introduce additional unit vectors in order to perform Fourier analysis in the gyrophase. Is there a method for computing the gyroaverages and gyroharmonics of a tensor in Cartesian coordinates without introducing more unit vectors than the necessary $b$?

A conceptually appealing way to answer this question is to first derive some coordinate-independent properties of the gyrosymmetry that would allow one to answer this question in an arbitrary coordinate system, and then specialize to cartesian coordinates. To our knowledge, this interesting mathematical exercise has not been discussed elsewhere in the literature, and so we will provide the details in the remainder of this section.

First notice that in a fibered coordinate system $(\xi_i,\theta)$ (this is shorthand for the sextuplet $(\xi_1,...,\xi_5,\theta)$), a function $f:U\rightarrow \mathbb{R}$ is gyrosymmetric if and only if
\begin{align}\label{rud_sym}
f(\xi_i,\theta+\psi)=f(\xi_i,\theta)
\end{align} 
for all constants $\psi$. If, for each $\psi\in\mathbb{R}~\text{mod}~2\pi$, we define the mapping $\Phi_\psi:U\rightarrow U$ using the formula
\begin{align}
\Phi_\psi(\xi_i,\theta)=(\xi_i,\theta+\psi),
\end{align}
then the condition given in Eq.\,(\ref{rud_sym}) can be re-expressed as
\begin{align}\label{fun}
\Phi_\psi^* f=f
\end{align}
for each $\psi\in\mathbb{R}~\text{mod}~2\pi$. Here $\Phi_\psi^*$ denotes the pullback operator on functions, $\Phi_\psi^*f=f\circ\Phi_\psi$.

While the formula (\ref{rud_sym}) only makes literal sense in a fibered coordinate system, the family of mappings $\Phi_\psi$ can actually be given a coordinate independent definition. Indeed, in cartesian position and velocity coordinates we have\footnote{In spite of its simplicity and utility, this formula only seems to have been noticed recently. It can be found in the literature in Ref. \onlinecite{BuQi12}. It was also independently discovered by Zhi Yu\cite{zhi}, but never published.}
\begin{align}\label{gyrosymmetry}
&\Phi_\psi(x,v)=\\
&(x,v\cdot b(x)b(x)+\cos(\psi) b(x)\times(v\times b(x))+\sin(\psi) v\times b(x)).\nonumber
\end{align}
Thus, gyrosymmetric functions $f:M\rightarrow \mathbb{R}$ can be alternately characterized as those functions that satisfy the analogue of Eq.\,(\ref{fun}), $\Phi_\psi^*f=f$ for each $\psi\in\mathbb{R}~\text{mod}~2\pi$. 

What about more general tensor fields? Because the pullback operator of a mapping $M\rightarrow M$ is well defined on the entire tensor algebra, it is tempting to postulate that a tensor field $\tau$ is gyrosymmetric if and only if $\Phi_\psi^*\tau=\tau$ for all $\psi\in\mathbb{R}~\text{mod}~2\pi$. This is indeed correct; it is a straightforward exercise to verify that this characterization is equivalent to the usual one stated in the previous section.

What is going on here? If we fix a $\psi\in\mathbb{R}~\text{mod}~2\pi$, then the mapping $\Phi_\psi:M\rightarrow M$ can be regarded as a global rearrangement, or relabeling, of points in $M$. If we regard $\Phi_\psi$ as pointing from the ``new arrangement" to the ``old arrangement", then $\Phi_\psi^*\tau$ is nothing more than $\tau$, regarded as a tensor in the old arrangement of $M$, expressed in the new arrangment. Thus, from this point of view, we see that gyrosymmetric tensors are precisely those tensors whose form is invariant under any of the rearrangments in the family $\Phi_\psi$.

With this coordinate-independent characterization of gyrosymmetric tensors in hand, we now seek a corresponding coordinate-independent version of Fourier analysis in the gyrophase $\theta$. The catch is that we do not desire to work with the gyrophase coordinate $\theta$ directly as the latter is only defined in fibered coordinate systems. Instead we will use the parameter $\psi$ in the family of maps $\Phi_\psi$ as a surrogate of sorts. 

Given an arbitrary tensor $\tau$, set $\tau_\psi=\Phi_\psi^*\tau$. $\tau_\psi$ can be regarded as a periodic tensor field-valued function of the single variable $\psi$ with period $2\pi$. Therefore it admits a Fourier expansion
\begin{align}\label{real_fourier}
\tau_\psi=\left<\tau\right>+\sum_{k=1}^\infty (\Pi_k\tau)\cos(\psi)+(\bar{\Pi}_k\tau)\sin(\psi),
\end{align} 
where the tensor fields $\left<\tau\right>$, $\Pi_k\tau$, and $\bar{\Pi}_kf$ are given by
\begin{align}\label{fourier_inversion}
\left<\tau\right>&=\frac{1}{2\pi}\int_0^{2\pi}(\Phi_\psi^*\tau)d\psi\\
\Pi_k\tau&=\frac{1}{\pi}\int_0^{2\pi}(\Phi_\psi^*\tau)\cos(k\psi)d\psi\nonumber\\
\bar{\Pi}_k\tau&=\frac{1}{\pi}\int_0^{2\pi}(\Phi_\psi^*\tau)\sin(k\psi)d\psi.\nonumber
\end{align}
Note that $\Pi_k\tau$ and $\bar{\Pi}_k\tau$ are not gyrosymmetric tensors. Instead they satisfy the identities
\begin{align}\label{fourier_symmetry}
\Phi_\psi^*(\Pi_k\tau)&=\cos(k\psi)(\Pi_k\tau)+\sin(k\psi)(\bar{\Pi}_k\tau)\\
\Phi_\psi^*(\bar{\Pi}_k\tau)&=-\sin(k\psi)(\Pi_k\tau)+\cos(k\psi)(\bar{\Pi}_k\tau).\nonumber
\end{align}
However, as the notation suggests, $\left<\tau\right>$ is indeed gyrosymmetric.

 The validity of these formulae only relies on the fact that the mapping $\Phi_\psi$ satisfies the group property $\Phi_{\psi_1+\psi_2}=\Phi_{\psi_1}\circ\Phi_{\psi_2}$. Therefore, they may be applied to tensors on any manifold equipped with such a family of mappings. These formulae also represent a shift in perspective from the Fourier analysis one would usually employ in fibered coordinate systems. To see this, consider a fibered coordinate system and the Fourier expansion of a function $f(\xi_i,\theta)=f_0(\xi_i)+\sum_k a_k(\xi_i)\cos(k\theta)+b_k(\xi_i)\sin(k\theta)$. It is not true that $\Pi_kf=a_k$; instead $(\Pi_kf)(\xi_i,\theta)=a_k(\xi_i)\cos(k\theta)+b_k(\xi_i)\sin(k\theta)$. Thus, the operators $\Pi_k$ and $\bar{\Pi}_k$ are not merely calculating the usual Fourier coefficients $a_k(\xi_i),b_k(\xi_i)$. Moreover, $\Pi_kf$ is a genuine scalar whereas the usual Fourier coefficients $a_k(\xi_i),b_k(\xi_i)$ have non-trivial transformation laws when passing from one fibered coordinate system to another (i.e. a change of gyrogauge). Indeed, if $(\xi_i,\theta)$ and $(\xi_i,\theta^\prime)$ are two fibered coordinate systems related by $\theta^\prime=\theta+\phi(\xi_i)$, then the usual Fourier coefficients in the primed coordinates $a_k^\prime(\xi_i),b_k^\prime(\xi_i)$ are related to the usual Fourier coefficients in the unprimed coordinates by
 \begin{align}
 a_k(\xi_i)&=a_k^\prime(\xi_i)\cos(k\phi(\xi_i))+b_k^\prime(\xi_i)\sin(k\phi(\xi_i))\\
 b_k(\xi_i)&=b_k^\prime(\xi_i)\cos(k\phi(\xi_i))-a_k^\prime(\xi_i)\sin(k\phi(\xi_i)).
 \end{align}
Therefore, the coordinate-independent Fourier analysis given by equations (\ref{real_fourier}) and (\ref{fourier_inversion}) calculate gyrogauge invariant combinations of the usual Fourier coefficients.

The Fourier inversion formula, Eq.\,(\ref{fourier_inversion}), together with the invariance properties given in Eq.\,(\ref{fourier_symmetry}), is sufficient to solve all of the linear partial differential equations that one encounters while deriving expressions for the $G_n(\epsilon)$ in any coordinate system. This is because: (a) all tensors encountered while deriving the guiding center expansion contain finitely many gyroharmonics; (b) the differential operator $L_{\xi}$, where $\xi=v\times b\cdot\frac{\partial}{\partial v}$ becomes an algebraic operator on gyroharmonics,
\begin{align}
L_{\xi}\left<\tau\right>&=0\\
L_{\xi}(\Pi_k\tau)&=k\bar{\Pi}_k\tau\\
L_{\xi}(\bar{\Pi}_k)&=-k\Pi_k\tau;
\end{align}
and (c) $L_\xi$ is the only partial differential operator that ever needs to be inverted. Thus, we have effectively solved the problem of performing ``Fourier analysis in $\theta$" in cartesian position and velocity coordinates without ever referring to fibered coordinate systems. We have incorporated this solution into our algorithm as
\\ \\
\emph{Resolution of Difficulty 3:}
Gyroaverages and gyroharmonics are calculated in cartesian position and velocity coordinates using Eqs.\,(\ref{fourier_inversion}) and (\ref{fourier_symmetry}).
\\ \\ 
\section{\label{five}The Algorithm}
As discussed in section \ref{two}, the goal of the algorithm is to find a transformation $T_\epsilon$ in the form given in Eq.\,\ref{gen_trans} such that $T_{\epsilon*}\mathbf{d}\vartheta_\epsilon$ and $T_{\epsilon*}H$ are each gyrosymmetric to all orders in $\epsilon$ (section \ref{three} gives the general definition of a gyrosymmetric tensor).
This $T_\epsilon$ consists of a concatenated sequence of transformations of the form $\exp(Y)$. Thus, we are free to think of $T_\epsilon$ as the result of many intermediate transformations, each closer to the identity transformation than the last. Our algorithm proceeds by finding expressions for these intermediate transformations (which amounts to specifying a $G_n(\epsilon)$), one at a time, according to the following recursive procedure.

Suppose that some finite number of intermediate transformations have been performed. Let $\Theta_\epsilon$ and $\mathcal{H}_\epsilon$ denote the resulting one-form and Hamiltonian following this partial rearrangement of $M$, and assume they have the form:
\begin{align}\label{standard_form}
\Theta_\epsilon=&\vartheta_0+\epsilon\vartheta_1+...+\epsilon^N\vartheta_N+\sum_{k=1}^\infty\epsilon^{N+k}\alpha_k\\
    \mathcal{H}_\epsilon=&H_0+...+\epsilon^{N-2}H_{N-2}+\sum_{k=1}^\infty \epsilon^{N-2+k}h_k,\nonumber
\end{align}
where $N>1$, the $\vartheta_j$ and $H_j$ are all gyrosymmetric, and the $\alpha_j$ and $h_j$ are not necessarily so. Suppose further that $\Xi_\epsilon=\vartheta_0+...+\epsilon^N\vartheta_N$ satisfies the three properties
\\ \\
ND1\footnote{ND stands for ``non-degenerate''.}: $\mathbf{d}\Xi_\epsilon$ is a non-degenerate two-form.
\\ \\
ND2:  If $\beta$ is an $\epsilon$-independent one-form, then the vector field $Y(\epsilon)$ defined by $\text{i}_{Y(\epsilon)}\mathbf{d}\Xi_\epsilon=\beta$ (i.e. $Y$ is the application of the Poisson tensor defined by $\Xi_\epsilon$ to $\beta$) is $O(\epsilon^{-2})$.
\\ \\
ND3: When $\beta=-\mathbf{d}H_0$, the leading order behavior of $Y(\epsilon)$ is given by $\frac{|B|}{\epsilon^2} \xi\equiv\frac{|B|}{\epsilon^2} v\times b\cdot\frac{\partial}{\partial v}$.
\\ \\
In this setting, which will serve as our inductive assumption, it is possible to find a transformation $\exp(-G(\epsilon))$, for some small vector field $G(\epsilon)$, such that after this transformation, the one-form and the Hamiltonian have the same form as in Eq.\,(\ref{standard_form}), but with $N$ replaced with $N+1$, i.e. the one-form and Hamiltonian are each gyrosymmetric to one higher order than previously. This also means that the transformed $\Xi_\epsilon$ will automatically continue to satisfy properties ND1-3. We will call a $G(\epsilon)$ that produces a transformation $\exp(G(\epsilon))$ with the latter two properties a \emph{recursive} vector field.

To see that one can in fact find \emph{many} recursive vector fields under the inductive assumption, let $G(\epsilon)$ be a vector field that solves the algebraic equation (see appendix \ref{appB} for a solution method)
\begin{align}\label{G}
\text{i}_{G(\epsilon)}\mathbf{d}\Xi_\epsilon+\epsilon^{N+1}\alpha_1+\epsilon^{N+1}\mathbf{d}S=\text{i}_{\left<G(\epsilon)\right>}\mathbf{d}\Xi_\epsilon+\epsilon^{N+1}\left<\alpha_1\right>,
\end{align}
where $S$ is the unique function with $\left<S\right>=0$ (see section \ref{four} for the definition of the general tensor gyroaverage operator $\left<\right>$) that solves the partial differential equation (see appendix \ref{appC} for a solution method)
\begin{align}\label{S}
h_1-|B|\text{i}_\xi\alpha_1-|B|\text{i}_\xi\mathbf{d}S=\left<h_1\right>-|B|\text{i}_\xi\left<\alpha_1\right>.
\end{align}
Note that the oscillatory part of $G(\epsilon)$, $\tilde{G}(\epsilon)=G(\epsilon)-\left<G(\epsilon)\right>$, is then uniquely determined, but the gyroaverage $\left<G(\epsilon)\right>$ is left completely free. Constrain the latter so that it satisfies
\begin{align}\label{freedom}
\text{i}_{\left<G(\epsilon)\right>}\mathbf{d}\Xi_\epsilon=\epsilon^{N+1}\gamma,
\end{align}  
where $\gamma$ is any $\epsilon$-independent gyrosymmetric one-form. Note that $\left<S\right>=0$ is not an arbitrary choice, but is necessary in order for Eq.\,(\ref{G}) to be self consistent. Indeed, upon gyroaveraging Eq.\,(\ref{G}), we see that $\mathbf{d}\left<S\right>=0$, which implies $\left<S\right>$ is constant. This constant is clearly inconsequential, and so we set it to zero. 

As is readily verified, such a $G(\epsilon)$ satisfies the following important properties.
\\ \\
P1: $G(\epsilon)=O(\epsilon^{N-1}),$ but will be a rational function of $\epsilon$.
\\ \\
P2: Upon applying the transformation $\exp(-G(\epsilon))$, the transformed one-form (modulo closed one-forms) and Hamiltonian, $\Theta_\epsilon^\prime$  and $\mathcal{H}_\epsilon^\prime$, become
\begin{align}
\Theta_\epsilon^\prime=&\vartheta_0+...+\epsilon^N\vartheta_N\\
                                    &+\epsilon^{N+1}\left(\left<\alpha_1\right>+\gamma\right)+O(\epsilon^{N+2}),\nonumber
\end{align}
and
\begin{align}
\mathcal{H}_\epsilon^\prime=&H_0+...+\epsilon^{N-2}H_{N-2}\\
                                               &+\epsilon^{N-1}\left(\left<h_1\right>+|B|\text{i}_\xi\gamma\right)+O(\epsilon^{N}).\nonumber
\end{align}
Thus, the entire family of $G(\epsilon)$ just defined, a family which may be regarded as being parameterized by the arbitrary gyrosymmetric one-form $\gamma$, consists of recursive vector fields. We refer the reader to Appendix \ref{appTrans} to clearly see the motivation for choosing Eqs.\,(\ref{G}) and (\ref{S}).

With these recursive vector fields in hand, all that we must now show is that there is some base case, consisting of a one-form and Hamiltonian in the form specified by Eq.\,(\ref{standard_form}), from which our recursive algorithm can start. Unfortunately this base case clearly cannot be the natural choice, $\Theta_\epsilon=\mathbf{A}\cdot dx+\epsilon v\cdot dx $ and $\mathcal{H}_\epsilon=\frac{1}{2}\epsilon^2v\cdot v$, as this pair is not in the form specified by Eq.\,(\ref{standard_form}). However, this issue is easy to resolve. First, notice that if the transformed $\frac{1}{\epsilon^2}H$ is gyrosymmetric, then $H$ will be too. Therefore, remove the $\epsilon^2$ from the Hamiltonian. Second, recognize that we are free to perform a preparatory transformation before finding the $G_n(\epsilon)$. 
In particular, we can apply a preparatory near-identity transformation of the form $\exp(-\epsilon G_0)$ that transforms $\mathcal{H}_\epsilon=\frac{1}{2}v\cdot v$ and $\Theta_\epsilon=\mathbf{A}\cdot dx+\epsilon v\cdot dx $ into the form specified by Eq.\,(\ref{standard_form}) with $N=2$. For instance, with
\begin{align}\label{prep}
&G_0=-\frac{v\times b}{|B|}\cdot\frac{\partial}{\partial x}\\
               &+\bigg(\frac{(v\cdot b)\nabla b\cdot(b\times v)}{|B|}+\frac{(v\times b)\cdot\nabla b\cdot vb}{2|B|}\nonumber\\
               &~~~~~~+\frac{(v\cdot b)(b\cdot\nabla\times b)b\times(v\times b)}{|B|}\bigg)\cdot\frac{\partial}{\partial v}\nonumber
\end{align}
then we arrive at the satisfactory starting point
\begin{align}\label{standard_one_form}
&\Theta_\epsilon=\mathbf{A}\cdot dx+\epsilon (v\cdot b)b\cdot dx\\
                          &+\frac{1}{2|B|}\epsilon^2\bigg(v\times b\cdot dv-(v\cdot b)[\nabla b\cdot (v\times b)]\cdot dx\bigg)+O(\epsilon^3)\nonumber\\
                          &\mathcal{H}_\epsilon=\frac{1}{2}v\cdot v+O(\epsilon),\nonumber
\end{align}
which can be verified by directly calculating Lie derivatives (using \emph{VEST}, for instance). More generally, $G_0$ must be chosen so that, after applying the preparatory transformation generated by $G_0$, the one-form has the form $\Theta_\epsilon=\theta_0+\epsilon\theta_1+\epsilon^2\theta_2+O(\epsilon)^2$ for gyrosymmetric one-forms $\theta_0,\theta_1,\theta_2$, and $\mathbf{d}\Theta_\epsilon$ is non-degenerate. After applying such a preparatory transformation, properties ND1-3 will automatically be satisfied, and the inductive procedure can begin. We found the $G_0$ just given, as well as the $G_0$ given in the second example below by directly analyzing the transformed one-form and demanding that it satisfy these properties. On the other hand, it would be interesting to find the most general $G_0$ that accomplishes the preparatory transformation. We leave this to future studies.

To summarize, our algorithm for finding the $G_n(\epsilon)$ that generate $T_\epsilon$ proceeds as follows. 
\\ \\
\emph{1:} Because any real calculation can only calculate a finite number of the $G_n(\epsilon)$, when one stops calculating additional $G_n(\epsilon)$, the one-form and Hamiltonian will be of the form specified in Eq.\,(\ref{standard_form}) with $N=N_{\text{max}}$. Therefore, specify the desired $N_{\text{max}}$.  
\\ \\
\emph{2:} Define three integers $N$, $M$, and $l$ with initializations $N=2$, $M=N_{\text{max}}-2$, and $l=1$. 
\\ \\
\emph{3:} Apply a preparatory transformation, such as that given in Eq.\,(\ref{prep}), so that the one-form and Hamiltonian have the form specified in Eq.\,(\ref{standard_form}). Only the first $M$ non-gyrosymmetric terms must be calculated in each case.
\\ \\
\emph{4:} Find a recursive vector field $G_l(\epsilon)$. Use Eq.\,(\ref{G}) to find the unique oscillatory part $\tilde{G}_l(\epsilon)$, and specify the gyroaveraged part $\left<G_l(\epsilon)\right>$ using an arbitrary gyrosymmetric one-form $\gamma_l$ according to Eq.\,(\ref{freedom}).
\\ \\
\emph{5:} Store $G_l(\epsilon)$. Set $l=l+1$, $N=N+1$, and $M=M-1$. 
\\ \\
\emph{6:} Using the recursive vector field just derived to specify the transformation, express the new one-form and Hamiltonian in the form specified in Eq.\,(\ref{standard_form}). Only the first $M$ non-gyrosymmetric terms must be calculated in each case. This amounts to applying $\exp(\text{i}_{G_l(\epsilon)}\mathbf{d})$ and $\exp(L_{G_l(\epsilon)})$ to the old one-form and Hamiltonian to generate the new $\alpha_i$ and $h_i$.
\\ \\
\emph{7:} If $N=N_{\text{max}}$, stop. Else, return to step \emph{4}.
\\ 

Note that different representations of the guiding center expansion will be generated for each choice of the sequence of gyrosymmetric one-forms $\gamma_l$ and the preparatory transformation generated by $G_0$. In particular, if one does not attempt to constrain the form of the transformed Hamiltonian, the $O(\epsilon^3)$ contribution to the transformed one-form can be any gyrosymmetric one-form whatsoever, including $0$. Likewise, if one does not attempt to constrain the form of the transformed one-form, then the $O(\epsilon)$ contribution to the transformed Hamiltonian can be specified arbitrarily (at least away from those points in phase space where $v\times b=0$ as $\text{i}_\xi\gamma=0$ at those points). 

Also note that the presence of a preparatory transformation implies that the complete transformation from the old phase space to the new, deformed phase space is given by $T_\epsilon\circ\exp(-\epsilon G_0)$, with $T_\epsilon=...\circ\exp(-G_2(\epsilon))\circ\exp(-G_1(\epsilon))$. In particular this motivates our convention for indexing the $G_n(\epsilon)$; by specifying the preparatory transformation as $\exp(-\epsilon G_0)$, we obtain the appealing result that $G_n(\epsilon)=O(\epsilon^n)$ in spite of the fact that both the preparatory transformation and the first transformation generated by the recursive procedure are $O(\epsilon)$.

Finally, note that now we can identity the precise manner in which this algorithm circumvents the order-mixing issue. The solution is the combination of the preparatory transformation and the fact the the recursive procedure determines each non-preparatory $G_n(\epsilon)$ without any knowledge of the form of $G_m(\epsilon)$ whenever $m>n$. The preparatory transformation is necessary to ensure that the equation defining $G_1(\epsilon)$, Eq.\,(\ref{G}), in the first recursive step can be solved. In particular, it guarantees that $\mathbf{d}(\vartheta_0+\epsilon\vartheta_1+\epsilon^2\vartheta_2)$ is invertible. After this initial step is complete, the two-form inversion required to compute the higher $G_n(\epsilon)$ is always possible because the two form in question is always a small perturbation to a two-form that is known to be invertible. Moreover, none of the quantities that appear in the equation defining $G_n(\epsilon)$, Eq.\,(\ref{G}), depend on knowledge of any of the $G_m(\epsilon)$ with $m>n$. The key observation that lead to this approach to determining the $G_n(\epsilon)$ was that it is not necessary to assume the $\epsilon$-dependence of each $G_n(\epsilon)$ \emph{a priori}. Indeed, the $G_n(\epsilon)$ determined by our scheme are ratios of polynomials in $\epsilon$, whereas the usual approach assumes each $G_n(\epsilon)$ is a monomial $\epsilon$.     
 
\section{\label{six}Two New Representations}
To illustrate the use of our algorithm, to suggest the relative ease of computing higher-order corrections to the guiding center expansion on a computer, and to emphasize the fact that there are still unexplored representations of the guiding center dynamics, we now turn to presenting the results of using the algorithm to derive two previously unexplored representations of the guiding center dynamics. Each representation we present here will choose $\gamma_l+\left<\alpha_1\right>=0$ in step \emph{4}, meaning each representation is closely related to the so-called Hamiltonian representation discussed in Ref. \onlinecite{BrTr}; this choice of the $\gamma_l$ leads to a transformed one-form that truncates at second order in $\epsilon$. While this property does not completely characterize the Hamiltonian representation of Ref. \onlinecite{BrTr}, it is a characteristic thereof. The two representations will differ from eachother in which preparatory transformation is used in step \emph{3} of the algorithm. Thus, these examples give a taste of the consequences of choosing different preparatory transformations, but not of different schemes for choosing $\gamma_l$. We will not evaluate either representation in terms of simplicity or time-validity, but instead leave this study to future work.

Before we present these representations, it is important that we point out a subtle aspect of our approach that follows from the fact that we work in cartesian position and velocity coordinates. By employing these coordinates, all of our calculations are done in the full six-dimensional single-particle phase space. Thus, the transformed one-form and Hamiltonian, $\Theta$ and $\mathcal{H}$, specified by our algorithm are quantities defined on a six-dimensional space. This implies that the new dynamical vector field (which specifies the equations of motion) specifies the evolution of the six variables $(X,Y,Z,v_x,v_y,v_z)$ (note that x=(X,Y,Z) in our notation). In particular, $\mu$, the magnetic moment is not a coordinate as it is in many treatments of guiding center theory. It is important to understand that in spite of this fact, these transformed six-dimensional equations nevertheless possess an \emph{exact} conservation law for any truncation of $\Theta$ and $\mathcal{H}$. This follows from the fact that the full six-dimensional transformed equations of motion are Hamiltonian, meaning they satisfy Hamilton's equations on the six-dimensional phase space $\text{i}_X\mathbf{d}\Theta=-\mathbf{d}\mathcal{H}$, where $X=(\dot{x},\dot{v})$. Thus, the Hamiltonian version of the Noether theorem implies that the function $\mu=\text{i}_\xi\Theta$ ($\xi=v\times b\cdot\frac{\partial}{\partial v}$ is the infinitesimal generator of the gyrosymmetry) is constant along trajectories of the vector field $X$ as a result of the invariance of $\Theta$ and $\mathcal{H}$ under the continuous family of symmetries defined by the gyrosymmetry $\Phi_\psi$. It is also important to realize that it is simple to obtain an expression of results produced by our algorithm in a coordinate system that uses the conserved quantity as a coordinate by using the six functions $(X,Y,Z,v_\parallel,\mu,\theta)$, with $\theta$ a gyrophase function defined relative to some perpendicular unit vectors, as coordinates. In this type of coordinate system, it will also be obvious how the one-form and Hamiltonian descend to the reduced (four-dimensional) phase space parameterized by $(X,Y,Z,v_\parallel)$ for fixed $\mu=\text{i}_\xi\Theta$. To illustrate this fact, we will present the one-form in each representation both in cartesian coordinates, as it is given directly by Mathematica, and in $(X,Y,Z,v_\parallel,\mu,\theta)$ coordinates, which we calculate by hand starting from the cartesian result. We will not do the same for the Hamiltonian as it will be trivial to express the Cartesian Hamiltonian in terms of $(X,Y,Z,v_\parallel)$.

For each representation, we will provide explicit expressions for the transformed one-form accurate to all orders in $\epsilon$. For the transformed Hamiltonians, we will provide $H_0$, $H_1$, and $H_2$. This information is enough to accurately express the full transformed equations of motion up to and including terms of order $\epsilon^2$. Moreover, the reduced equations of motion, which describe the evolution of the guiding center position $x$ and parallel velocity $v_\parallel$, can be specified up to and including terms of order $\epsilon^3$. For the sake of brevity, we will not display the $G_n(\epsilon)$. However, we stress that, when equipped with a copy of our code, finding these vector fields that specify the transformation would be a simple task for any interested reader. In fact, each of the results below takes about fifteen minutes to derive on a laptop computer equipped with \emph{VEST}. We would also like to stress that all of the following results have been checked thoroughly. Using \emph{VEST}, we have explicitly expanded the pushforward operator $T_{\epsilon*}$ into a series of Lie derivatives along the $G_n({\epsilon})$ and applied it to the zero'th order Hamiltonian and one-form. The resulting expressions agree exactly with the results reported below up to the relevant order. They have also been checked indirectly by verifying that we reproduce the well-known first-order correction to the magnetic moment adiabatic invariant (in untransformed variables), $\mu_1$ (see Ref. \onlinecite{WB}, for example): 
\begin{align}
\mu_1&=\frac{1}{|B|^2}\bigg(\frac{1}{4}v\cdot\nabla b\cdot (v\times b)(v\cdot b)\\
&-\frac{3}{4}(v\times b)\cdot\nabla b\cdot v (v\cdot b)-\frac{5}{4}b\times
\kappa\cdot v(v\cdot b)^2\nonumber\\
&+\frac{1}{2|B|}(v\times b)\cdot(v\times b)\nabla|B|\times b\cdot v\bigg).\nonumber
\end{align}
Moreover, we have investigated $\mu_2$ using \emph{VEST}. We have verified that the $\mu_2$ predicted by the $G_n(\epsilon)$ in each representation agree with one another. We have also directly verified that the time derivative of $\mu_0+\epsilon\mu_1+\epsilon^2\mu_2$ along the Lorentz force equations of motion $\dot{v}=|B|v\times b,\dot{x}=\epsilon v$ is $O(\epsilon^3)$ (in general, the adiabatic invariant series must satisfy $\frac{d}{dt}(\sum_{k=0}^N\epsilon^k\mu_k)=O(\epsilon^{N+1})$). Finally, we have compared our expression for $\mu_2$ with that given in Ref. \onlinecite{WB}, which is the only published $\mu_2$ applicable to general magnetic geometry we are aware of. Interestingly, our expression is numerically different from the result reported in Ref. \onlinecite{WB}. However, this difference is most likely due to a likely typographical error in Ref. \onlinecite{WB}, as explained in Appendix \ref{app_mu}. We also present our expression for $\mu_2$ in Appendix \ref{app_mu}. 
\\ \\
\emph{Example 1:}
\\ \\
This representation will be defined by the use of the preparatory transformation already given in Eq.\,(\ref{prep}) and by always choosing $\gamma_l+\left<\alpha_1\right>=0$. The consequences of these choices come in the form of a transformed one-form equal to that given in Eq.\,(\ref{standard_one_form}) to all orders, thus simplifying the form of the transformed Poisson bracket. In fact, by expressing Eq.\,(\ref{standard_one_form}) in the usual sort of fibered coordinate system as well as cartesian position and velocity coordinates, we see
\begin{align}
\Theta_\epsilon
=&\mathbf{A}\cdot dx+\epsilon (v\cdot b)b\cdot dx\\
                          &+\frac{1}{2|B|}\epsilon^2\bigg(v\times b\cdot dv-(v\cdot b)[\nabla b\cdot (v\times b)]\cdot dx\bigg)\nonumber\\
                          =&\mathbf{A}\cdot dx+\epsilon v_\parallel b\cdot dx+\epsilon^2\mu[d\theta-\mathbf{R}\cdot dx]\nonumber
\end{align}
where $\mu=\text{i}_\xi\Theta=\frac{v_\perp^2}{2|B|}$ and $\mathbf{R}=(\nabla e_1)\cdot e_2$, meaning the transformed Poisson bracket is exactly the same as that given in Ref. \onlinecite{GGI}. Meanwhile the transformed Hamiltonian is given by
\begin{align}
H_0=&\frac{1}{2}v\cdot v\\
H_1=&\frac{1}{2}(v\cdot b)\mu \tau\nonumber\\
H_2=&\mu^2\bigg(\frac{15}{16}(\nabla\cdot b)^2+\frac{1}{16}\kappa\cdot\kappa+\frac{1}{4}b\cdot\nabla(\nabla\cdot b)\nonumber\\
&~~-\frac{1}{16}\text{tr}[\nabla b\cdot \nabla b+\nabla b\cdot(\nabla b)^T]\nonumber\\
&~~-\frac{3}{4}\nabla\ln|B|\cdot\nabla\ln|B|+\frac{1}{4}\kappa\cdot\nabla\ln|B|+\frac{1}{4|B|}\nabla^2|B|\bigg)\nonumber\\
+\mu&\frac{(v\cdot b)^2}{|B|}\bigg(\frac{1}{8}\text{tr}[3\nabla b\cdot \nabla b-\nabla b\cdot(\nabla b)^T]+\frac{1}{8}(\nabla\cdot b)^2\nonumber\\
&~~~~~~~~+\frac{1}{2}b\cdot\nabla(\nabla\cdot b)+\frac{13}{8}\kappa\cdot\kappa-\frac{3}{2}\kappa\cdot\nabla\ln|B|\bigg)\nonumber\\
&-\frac{(v\cdot b)^4}{|B|^2}\bigg(\frac{1}{2}\kappa\cdot\kappa\bigg)\nonumber
\end{align}
where $\mu=\text{i}_\xi\Theta=\frac{(v\times b)\cdot(v\times b)}{2|B|}$, $\tau=b\cdot\nabla\times b$, and $\kappa=b\cdot\nabla b$. Note that the $H_2$ in this representation differs from the $H_2$ in Brizard and Tronko's Hamiltonian representation \cite{BrTr} (also see Ref. \onlinecite{Brizard_thesis}), although it is similarly complicated. This difference is not just apparent; we have rigorously compared our $H_2$ with Brizard and Tronko's using \emph{VEST} and found they are not numerically equal. In order to recover Brizard and Tronko's representation, it would be necessary to either: (a) choose $\left<\alpha_1\right>+\gamma_l=\mathbf{d}f_l$ with appropriately chosen $f_l$ (we have chosen $f_l=0$ in this example); (b) alter the preparatory transformation Eq.\,(\ref{prep}); or (c) do both (a) and (b).
\\ \\
\emph{Example 2:}
\\ \\
The second representation will also make the choice $\gamma_l+\left<\alpha_1\right>=0$, but the preparatory transformation $\exp(-\epsilon G_0)$ will be specified by
\begin{align}
&G_0=-\frac{v\times b}{|B|}\cdot\frac{\partial}{\partial x}+\frac{1}{|B|}\bigg((v\cdot b)(v\times b)\cdot\nabla b\\
&-2(v\cdot b)\nabla b\cdot(v\times b)+\frac{1}{4}v\cdot[\nabla b+\nabla b^T]\cdot(v\times b)b\nonumber\\
&~~~~~~~~~~~~~~~~~~~~~~~~~~~~+\frac{3}{4}b\cdot\kappa\times v (v\cdot b)b\bigg)\cdot\frac{\partial}{\partial v}\nonumber
\end{align}
This implies that the transformed one-form is given by
\begin{align}
&\Theta_\epsilon=\mathbf{A}\cdot dx+\epsilon(v\cdot b)b\cdot dx\\
&+\frac{1}{2|B|}\epsilon^2\bigg(v\times b\cdot dv-(v\cdot b)[\nabla b\cdot (v\times b)]\cdot dx-\mu |B|\tau b\cdot dx\bigg)\nonumber\\
&~~~=\mathbf{A}\cdot dx+\epsilon v_\parallel b\cdot dx+\epsilon^2\mu[d\theta-(\mathbf{R}+\frac{1}{2}\tau b)\cdot dx].\nonumber
\end{align}
Thus, the transformed Poisson bracket is the same as that given in Ref. \onlinecite{Brizard_thesis}. The transformed Hamiltonian is given by
\begin{align}
H_0&=\frac{1}{2}v\cdot v\\
H_1&=0\\
H_2&=\mu^2\bigg(\frac{15}{16}(\nabla\cdot b)^2+\frac{3}{16}\kappa\cdot\kappa+\frac{1}{4}b\cdot\nabla(\nabla\cdot b)\nonumber\\
&~~+\frac{1}{16}\text{tr}[\nabla b\cdot \nabla b-3\nabla b\cdot(\nabla b)^T]\nonumber\\
&~~-\frac{3}{4}\nabla\ln|B|\cdot\nabla\ln|B|+\frac{1}{4}\kappa\cdot\nabla\ln|B|+\frac{1}{4|B|}\nabla^2|B|\bigg)\nonumber\\
+\mu&\frac{(v\cdot b)^2}{|B|}\bigg(\frac{1}{8}\text{tr}[3\nabla b\cdot \nabla b-\nabla b\cdot(\nabla b)^T]+\frac{1}{8}(\nabla\cdot b)^2\nonumber\\
&~~~~~~~~+\frac{1}{2}b\cdot\nabla(\nabla\cdot b)+\frac{13}{8}\kappa\cdot\kappa-\frac{3}{2}\kappa\cdot\nabla\ln|B|\bigg)\nonumber\\
&-\frac{(v\cdot b)^4}{|B|^2}\bigg(\frac{1}{2}\kappa\cdot\kappa\bigg).\nonumber
\end{align}
Although the one-form, $H_0$, and $H_1$ in this example are the same as those given by Parra and Calvo\cite{PnC}, $H_2$ is in fact different. Using \emph{VEST}, we have found that Parra and Calvo's $H_2$ has a numerically different value than the $H_2$ in this example. 
\section{\label{seven}Conclusion}
We have reported, for the first time, on the automatic calculation of the guiding center expansion using a computer. In particular, we have implemented a novel Lie transform-based algorithm using the newly-developed Mathematica package \emph{VEST}\cite{SqBuQi13} and used it to derive two new representations of the guiding center equations of motion to the order relevant for studying issues related to the physics of toroidal momentum conservation in tokamaks. By proceeding in this manner, we have avoided the pitfalls associated with hand-made algebra errors and slashed the time required to perform the calculations from weeks to minutes. Readers interested in obtaining the Mathematica notebook we used to carry out our calculation can contact J. Squire via email at jsquire@princeton.edu.

There are a number of opportunities for extending this work. Because our algorithm provides the necessary tools to explore many representations of the guiding center expansion, it may be interesting to begin searching through different representations to find those with desirable properties such as simple transformed equations of motion. Likewise, it may be interesting to examine the time-validity of the transformed equations of motion in these different representations to see if some are worse than others. Kruskal's theory\cite{kruskal62} guarantees $1/\epsilon$ time-validity in all representations, but there may be representations that can do better. Yet another suitable application of our code would be pushing the calculation to higher order than previously calculated. For instance, it would be interesting to find $\mu_3$ and $H_3$.  Finally, there should be no great difficulty in extending both our algorithm and our implementation in Mathematica using \emph{VEST} to treat the gyrocenter transformation theory that forms the backbone of modern gyrokinetic theory.

% If in two-column mode, this environment will change to single-column format so that long equations can be displayed. 
%\begin{widetext}
%$$\mbox{put long equation here}$$
%\end{widetext}

% Figures should be put into the text as floats. 
% Use the graphics or graphicx packages (distributed with LaTeX2e).
% Put the label that you will use with \ref{} command in the braces of the \label{} command.
%
% \begin{figure}
% \includegraphics{image2.eps}%
% \end{figure}

% \begin{table}
% \end{tabular}
% \end{table}

\begin{acknowledgments}
% Put your acknowledgments here.
The authors would like to express their gratitude to B. Faber for his help in editing this manuscript. This work was supported by the U.S. Department of Energy under contract DE-AC02-09CH11466. 
\end{acknowledgments}

\appendix
\section{\label{appLie} Elements of Exterior Calculus}
In this appendix we will first list the basic identities commonly used when performing calculations with the exterior calculus. Then we will give the component-form of the basic operators $\mathbf{d}$, $L_Y$, $\text{i}_Y$ in the velocity phase space. For a much more thorough treatment of this topic, refer to Ref. \onlinecite{FoM}.

Let $\alpha_k$ and $\beta_l$ be $k$- and $l$-forms on the manifold $M$, respectively. Let $Y$ and $Z$ be vector fields on $M$. Then the following identities hold
\begin{align}
\alpha_k\wedge\beta_l&=(-1)^{kl}\beta_l\wedge\alpha_k\\
\text{i}_Y(\alpha_k\wedge\beta_l)&=(\text{i}_Y\alpha_k)\wedge\beta_l+(-1)^{k}\alpha_k\wedge(\text{i}_Y\beta_l)\\
\mathbf{d}(\alpha_k\wedge\beta_l)&=(\mathbf{d}\alpha_k)\wedge\beta_l+(-1)^k\alpha_k\wedge(\mathbf{d}\beta_l)\\
L_Y(\alpha_k\wedge\beta_l)&=(L_Y\alpha_k)\wedge\beta_l+\alpha_k\wedge(L_Y\beta_l)\\
\text{i}_Y\text{i}_Z&=-\text{i}_Z\text{i}_Y\\
L_Y&=\mathbf{d}\text{i}_Y+\text{i}_Y\mathbf{d}\label{cartan}\\
\mathbf{d}L_Y&=L_Y\mathbf{d}\\
\mathbf{d}\mathbf{d}&=0.
\end{align}
Let $F:M\rightarrow M$ and $\Phi:M\rightarrow M$ be smooth mappings with a smooth inverses $F^{-1}$ and $\Phi^{-1}$. One example of this sort of mapping from the main text is $\Phi_\psi$, for fixed $\psi$, whose inverse is $\Phi_{-\psi}$. The exterior calculus operations behave very well with respect to mappings. We will summarize this fact with a second list of identities. 
\begin{align}
F^*\Phi^*&=(\Phi\circ F)^*\\
F^*(\alpha_k\wedge\beta_l)&=(F^*\alpha_k)\wedge(F^*\beta_l)\\
F^*(\text{i}_Y\alpha_k)&=\text{i}_{F^*Y}(F^*\alpha_k)\\
F^*(L_Y\alpha_k)&=L_{F^*Y}(F^*\alpha_k)\\
F^*(\mathbf{d}\alpha_k)&=\mathbf{d}(F^*\alpha_k).
\end{align}
When $F=\exp(Y(\epsilon))$, with $Y(\epsilon)$ a vector field that tends to zero as $\epsilon\rightarrow 0$, we also have the asymptotic identities
\begin{align}
\exp(-Y(\epsilon))_*\tau&=\exp(Y(\epsilon))^*\tau\\
\exp(Y(\epsilon))^*\tau&=\tau+L_{Y(\epsilon)}\tau+\frac{1}{2}L^2_{Y(\epsilon)}T+...,
\end{align}
where $\tau$ is an arbitrary tensor.

The identities provided thus far, together with the fact that the wedge product is associative, are sufficient to verify all of the coordinate-independent manipulations of differential forms in the main text. In order to perform exterior calculus using \emph{VEST} it is also useful to have component-based expressions for the operators $\mathbf{d}$, $\text{i}_Y$, and $L_Y$. Actually, the relevant operators for the sake of performing the guiding center calculation are $\mathbf{d}$ on functions, $\text{i}_Y$ on one-forms, and $\text{i}_Y\mathbf{d}$ on both functions and one-forms.

Let $Y=Y^{xi}\frac{\partial}{\partial x^i}+Y^{vi}\frac{\partial}{\partial v^i}$, where the indices are summed from $i=1$ to $i=3$ (although we do so by habit, there is really no need to distinguish between covariant and contravariant indices in cartesian coordinates). Similarly, let $\alpha=\alpha_{xi}dx^i+\alpha_{vi}dv^i$. Denote derivatives of a scalar $f$ with respect to the $i$'th spatial argument and the $i$'th velocity argument with $f_{,i}$ and $f_{;j}$, respectively (note that $;$ does \emph{not} denote a covariant derivative). Then we have
\begin{align}
\mathbf{d}f&=f_{,i}dx^i+f_{;i}dv^i\\
\text{i}_Y\alpha&=\alpha_{xi}Y^{xi}+\alpha_{vi}Y^{vi}\\
\text{i}_Y\mathbf{d}f&=f_{,i}Y^{xi}+f_{;i}Y^{vi}\\
\text{i}_Y\mathbf{d}\alpha&=(\alpha_{xi,j}-\alpha_{xj,i})Y^{xj}dx^i\\
                                         &+(\alpha_{xi;j}Y^{vj}-\alpha_{vj,i}Y^{xj})dx^i\nonumber\\
                                         &+(\alpha_{vi,j}Y^{xj}-\alpha_{xj;i}Y^{xj})dv^i\nonumber\\
                                         &+(\alpha_{vi;j}-\alpha_{vj;i})Y^{vj}dv^i.\nonumber
\end{align}
Note that, by the identity given in Eq.\,(\ref{cartan}), the Lie derivative of a one-form, $L_Y\alpha=\mathbf{d}\text{i}_Y\alpha+\text{i}_Y\mathbf{d}\alpha$, can also be calculated using these component expressions.

\section{\label{appA} The Origin of Many Representations of The Guiding Center Expansion}
Suppose that a near-identity rearrangement of the phase space $T_\epsilon:M\rightarrow M$ is found that renders the transformed Lorentz vector field $X_H^\prime=T_{\epsilon*}X_H$ gyrosymmetric. As explained by Kruskal\cite{kruskal62}, at least one such transformation can be found using perturbation theory. In fact, as soon as one transformation in found, many more can be generated easily. This implies that there is much freedom in choosing $T_\epsilon$; each choice leads to a different representation of the guiding center equations of motion in the sense that $X_H^\prime$ will be different in each case. In this appendix we will explain the origin of this freedom by completely characterizing it.

First note that if $F:M\rightarrow M$ is a rearrangement of phase space (not necessarily near-identity) that commutes with the family of rearrangements $\Phi_\psi$ that define the gyrosymmetry (see section \ref{four}), i.e. $F\circ \Phi_\psi=\Phi_\psi\circ F$ for each $\psi\in\mathbb{R}~\text{mod}~2\pi$, then $X_H^{\prime\prime}\equiv F_*X_H^\prime$ is also gyrosymmetric. Indeed, 
\begin{align}
\Phi_\psi^*X_H^{\prime\prime}&=\Phi_{-\psi*}F_*X_H^\prime\\
                                                &=(\Phi_{-\psi}\circ F)_*X_H^\prime\nonumber\\
                                                &=(F\circ\Phi_{-\psi})_*X_H^\prime\nonumber\\
                                                &=F_*\Phi_\psi^* X_H^\prime=X_H^{\prime\prime},\nonumber
\end{align} 
where we have used the fact that $X_H^\prime$ is gyrosymmetric and $F\circ\Phi_\psi=\Phi_\psi\circ F$. This tells us that, given one of the near-identity rearrangements of phase space guaranteed by Kruskal, we can find many more by following the latter with any near-identity rearrangement of phase space that commutes with $\Phi_\theta$.

In fact \emph{all} of the rearrangements that fit into Kruskal's theory can be found in this manner. To be precise, suppose that $R_\epsilon:M\rightarrow M$ and $Q_\epsilon:M\rightarrow M$ are two near identity rearrangements that render $X_H$ gyrosymmetric, so that they fit into Kruskal's theory. Then, by definition, $\Phi_\psi^*(R_{\epsilon*}X_H)=R_{\epsilon*}X_H$ and $\Phi_\psi^*(Q_{\epsilon*}X_H)=Q_{\epsilon*}X_H$. Thus, both $Q_\epsilon$ and $R_\epsilon$ define symmetry transformations on the original arrangement of phase space, $\bar{\Phi}^Q_\psi=Q_\epsilon^{-1}\circ\Phi_\psi\circ Q_\epsilon$ and $\bar{\Phi}^R_\psi=R_\epsilon^{-1}\circ\Phi_\psi\circ R_\epsilon$ that leave $X_H$ invariant. Kruskal has proven in Ref. \onlinecite{kruskal62} that these two symmetry transformations are in fact identical to all orders in $\epsilon$; $\bar{\Phi}_\psi\equiv\bar{\Phi}^Q_\psi=\bar{\Phi}^R_\psi$. It follows then that the rearrangement $F=Q_\epsilon\circ R_\epsilon^{-1}$ commutes with $\Phi_\psi$. But this is precisely the rearrangement that sends $R_{\epsilon*}X_H$ into $Q_{\epsilon*}$, which tells us that each representation of the guiding center equations of motion can be reached from a given one by applying a near-identity transformation that commutes with $\Phi_\theta$. 

Note that if a rearrangement of the form $\exp(Y)$, for some vector field $Y$, commutes with $\Phi_\theta$, then it must be true that $Y=\left<Y\right>$. This is why one should expect complete freedom to choose the $\left<G_n(\epsilon)\right>$ for each $G_n(\epsilon)$ appearing in the Lie transform ansatz given in Eq.\,(\ref{gen_trans}).

\section{\label{appB} A General Formula For Inverting Exact Lagrange Tensors Defined On The Velocity Phase Space}
Step \emph{4} in our algorithm involves solving an algebraic equation of the form $\text{i}_X\mathbf{d}\Xi=-\beta$ for the vector field $X$ given a non-degenerate two-form $\mathbf{d}\Xi$ and an arbitrary one-form $\beta$. In this appendix we will present explicit expressions for the components of $X=X^{xi}\frac{\partial}{\partial x^i}+X^{vi}\frac{\partial}{\partial v^i}$ in terms of the components of $\Xi=A_idx^i+B_idv^i$ and $\beta=\beta_{xi}dx^i+\beta_{vi}dv^i$.

We proceed by making use of the linear isomorphism between the space of vector fields and the space of five-forms induced by the Liouville volume form
\begin{align}
\Omega=&\frac{1}{6}\mathbf{d}\Xi\wedge\mathbf{d}\Xi\wedge\mathbf{d}\Xi.
\end{align}
This isomorphism is given by $X\mapsto\text{i}_X\Omega$. One can easily prove that this is an isomorphism using the fact that the non-degeneracy of $\mathbf{d}\Xi$ implies that $\Omega$ is nowhere vanishing. The reason this isomorphism is useful is that it is easier to find $\mathcal{X}\equiv\text{i}_X\Omega$ than $X$. Indeed, upon wedge multiplying $\mathbf{d}\Xi\wedge\mathbf{d}\Xi$ into both sides of the equation $\text{i}_X\mathbf{d}\Xi=-\beta$, we obtain
\begin{align}
\mathcal{X}=-\frac{1}{2}\mathbf{d}\Xi\wedge\mathbf{d}\Xi\wedge\beta.
\end{align} 
By explicitly calculating the right hand side of the last expression in components, and then inverting the isomorphism $X\mapsto\text{i}_X\Omega$, we obtain
\begin{align}
X^{xn}=&-\frac{1}{\mathcal{D}}\bigg(\epsilon^{ilk}\epsilon^{mjn}\beta_{xm}B_{k;l}(B_{i,j}-A_{j;i})\\
            &~~~~~~~~~~~+\epsilon^{lkm}\epsilon^{jin}\beta_{vm}A_{i,j} B_{k;l}\nonumber\\
            &+\frac{1}{2}\epsilon^{kim}\epsilon^{jln}\beta_{vm}(B_{i,j}-A_{j;i})(B_{k,l}-A_{l;k})\bigg)\nonumber\\
X^{vn}=&\frac{1}{\mathcal{D}}\bigg(\epsilon^{jil}\epsilon^{kmn}\beta_{vm}A_{i;j}(B_{k,l}-A_{l;k})\\
            &~~~~~~~~~~~+\epsilon^{mji}\epsilon^{lkn}\beta_{xm}A_{i,j}B_{k;l}\nonumber\\
            &+\frac{1}{2}\epsilon^{mjl}\epsilon^{kin}\beta_{xm}(B_{i,j}-A_{j;i})(B_{k,l}-A_{l;k})\bigg),\nonumber            
\end{align}
where the function $\mathcal{D}$ is defined by the relation $\Omega=\mathcal{D}dx^1\wedge dx^2\wedge dx^3\wedge dv^1\wedge dv^2\wedge dv^3$.

\section{\label{appC} Solving The Equation for S}
In order to complete step \emph{4} in our algorithm, the partial differential equation
\begin{align}
h_1-|B|\text{i}_\xi\alpha_1-|B|\text{i}_\xi\mathbf{d}S=\left<h_1\right>-|B|\text{i}_\xi\left<\alpha_1\right>
\end{align}
must be solved for $S$ given $|B|$, $\alpha_1$, $h_1$, and the constraint $\left<S\right>=0$. As is readily verified by gyroaveraging the equation, an equivalent condition on $S$ is that it be chosen to eliminate the non-zero gyroharmonics of the quantity $h_1-|B|\text{i}_\xi\alpha_1-|B|\text{i}_\xi\mathbf{d}S$. 

Let $\nu=h_1-|B|\text{i}_\xi\alpha_1$. Using Eq.\,(\ref{fourier_inversion}), we see that the Fourier expansion of $\nu_\psi=\Phi_\psi^*\nu$ is given by
\begin{align} 
\nu_\psi=\left<\nu\right>+\sum_{k=1}^\infty \Pi_k\nu\cos(k\psi)+\bar{\Pi}_k\nu \sin(k\psi).
\end{align}
Using the identities $L_\xi\Pi_k S=k\bar{\Pi}_kS$ and $L_\xi\bar{\Pi}_k S=-k\Pi_k S$, we also see that the Fourier expansion of $\Phi_\psi^*|B|\text{i}_\xi\mathbf{d}S=|B|L_\xi S_\psi$ is given by
\begin{align}
|B|L_\xi S_\psi=\sum_{k=1}^\infty|B|k\bar{\Pi}_kS\cos(k\psi)-|B|k\Pi_kS\sin(k\psi).
\end{align} 
Therefore, $S$ must be given by
\begin{align}
S=&\sum_{k=1}^\infty\Pi_k S\\
   =&\sum_{k=1}^\infty-\frac{\bar{\Pi}_k\nu}{|B|k}.\nonumber
\end{align}

\section{\label{appTrans} Motivation for Equations Defining The $G_n(\epsilon)$}
In this appendix, we will motivate Eqs.\,(\ref{G}) and (\ref{S}).

Suppose the one-form and Hamiltonian are expressed in the form
\begin{align}%\label{standard_form}
\Theta_\epsilon=&\vartheta_0+\epsilon\vartheta_1+...+\epsilon^N\vartheta_N+\sum_{k=1}^\infty\epsilon^{N+k}\alpha_k\\
    \mathcal{H}_\epsilon=&H_0+...+\epsilon^{N-2}H_{N-2}+\sum_{k=1}^\infty \epsilon^{N-2+k}h_k,\nonumber
\end{align}
with each $\theta_i$ and $H_i$ gyrosymmetric and $N>1$. Define $\Xi=\vartheta_0+\epsilon\vartheta_1+...+\epsilon^N\vartheta_N$. Consider changing coordinates on the six-dimensional phase space using the mapping $\exp(-G(\epsilon))$ where $G(\epsilon)$ is some vector field that tends to zero as $\epsilon\rightarrow 0$. To be precise, consider $\exp(-G(\epsilon))$ as the map that sends old coordinates into new coordinates. Note that we have not specified the order of $G(\epsilon)$. Then the one-form and Hamiltonian expressed in the new coordinates are $\Theta_\epsilon^\prime=\exp(-G(\epsilon))_{*}\Theta_\epsilon$ and $\mathcal{H}_\epsilon^\prime=\exp(-G(\epsilon))_{*}\mathcal{H}_\epsilon$, respectively. Moreover, because $G(\epsilon)$ is small as $\epsilon\rightarrow 0$, we may expand these pushforward operators into a series of Lie derivatives, thereby obtaining the expressions
\begin{align}
\Theta_\epsilon^\prime&=\Xi+\text{i}_{G(\epsilon)}\mathbf{d}\Xi+\epsilon^{N+1}\alpha_1+\delta\Theta\\
\mathcal{H}_\epsilon^\prime&=H_0+...+\epsilon^{N-2}H_{N-2}+\text{i}_{G(\epsilon)}\mathbf{d}H_0+\epsilon^{N-1}h_1+\delta\mathcal{H},
\end{align}
where $\delta\Theta$ and $\delta\mathcal{H}$ have not been displayed for simplicity, but are not necessarily higher order in $\epsilon$.

These expressions for the transformed one-form and Hamiltonian are valid for any $G(\epsilon)$ that tends to zero as $\epsilon\rightarrow 0$. Assuming the non-degeneracy conditions  ND1-3 given in the main text, we will now use them to choose a specific $G(\epsilon)$ such that the transformed one-form and Hamiltonian are gyrosymmetric to one higher order than they were initially. 

First we demand that the one-form $\nu\equiv\text{i}_{G(\epsilon)}\mathbf{d}\Xi+\epsilon^{N+1}\alpha_1$ should be gyrosymmetric up to an $O(\epsilon^{N+1})$ exact differential. This implies $\nu+\epsilon^{N+1}\mathbf{d}S=\left<\nu\right>$, or
\begin{align}\label{aG}
\text{i}_{G(\epsilon)}\mathbf{d}\Xi+\epsilon^{N+1}\alpha_1+\epsilon^{N+1}\mathbf{d}S=\text{i}_{\left<G(\epsilon)\right>}\mathbf{d}\Xi+\epsilon^{N+1}\left<\alpha_1\right>.
\end{align} 
Note that this requirement forces the constraint $\left<\mathbf{d}S\right>=0$, as can be seen by gyroaveraging the previous expression. Also note that this expression is identical to Eq.\,\ref{G}. Finally, note that the gyroaverage of $G(\epsilon)$ completely disappears from the expression if we decompose $G(\epsilon)$ as $\left<G(\epsilon)\right>+\tilde{G}(\epsilon)$. Indeed, we have
\begin{align}\label{tildes}
\text{i}_{\tilde{G}(\epsilon)}\mathbf{d}\Xi=-\epsilon^{N+1}(\mathbf{d}\tilde{S}+\tilde{\alpha_1}).
\end{align}
Therefore, using the non-degeneracy conditions, we can impose the constraint $\text{i}_{\left<G(\epsilon)\right>}\mathbf{d}\Xi=\epsilon^{N+1}\gamma$, where $\gamma$ is an arbitrary $\epsilon$-independent gyrosymmetric one-form.

By condition ND1 and ND2, for a given $S$, Eq.\,(\ref{tildes}) has a unique $O(\epsilon^{N-1})$ solution for $\tilde{G}(\epsilon)$ given by inverting the two-form $\mathbf{d}\Xi$. Likewise, $\left<G(\epsilon)\right>=O(\epsilon^{N-1})$. Thus, $G(\epsilon)=O(\epsilon^{N-1})$. Using this result, it is straightforward to verify that, when $G(\epsilon)$ is chosen according to Eq.\,(\ref{aG}), $\delta\Theta=O(\epsilon^{N+2})$. Therefore, $\Theta_\epsilon^\prime$ is gyrosymmetric to one higher order in $\epsilon$ than it was before applying the coordinate transformation. Even more, this conclusion still holds for any choice of the function $S$.

This freedom in the selection of $S$ can be used to make the transformed Hamiltonian $\mathcal{H}_\epsilon^\prime$ gyrosymmetric to one higher order. To see this, first note that because $G(\epsilon)=O(\epsilon^{N-1})$, $\delta\mathcal{H}=O(\epsilon^{N})$. Thus, in order to make $\mathcal{H}_\epsilon^\prime$ gyrosymmetric to one higher order in $\epsilon$, all that we must do is ensure that the function $f=\text{i}_{G(\epsilon)}\mathbf{d}H_0+\epsilon^{N-1}h_1$ is gyrosymmetric at $O(\epsilon^{N-1})$. To see that this can be accomplished for a particular choice of $S$, define the vector field $X_{H_0}$ using the formula $\text{i}_{X_{H_0}}\mathbf{d}\Xi=-\mathbf{d}H_0$. This definition makes sense in light of condition ND1. In terms of $X_{H_0}$, $f$ can be written $f=\text{i}_{X_{H_0}}\text{i}_{G(\epsilon)}\mathbf{d}\Xi+\epsilon^{N-1}h_1$. Then, by Eq.\,(\ref{aG}), we have
\begin{align}
f=\epsilon^{N+1}\text{i}_{X_{H_0}}(\gamma-\tilde{\alpha}_1-\mathbf{d}S)+\epsilon^{N-1}h_1.
\end{align}
But by condition ND3, the leading-order contribution to $X_{H_0}$ is given by $\frac{|B|}{\epsilon^2} \xi$, which implies
\begin{align}
f=\epsilon^{N-1}|B|\text{i}_{\xi}(\gamma-\tilde{\alpha}_1-\mathbf{d}S)+\epsilon^{N-1}h_1+O(\epsilon^{N}).
\end{align}
Thus, the condition that all gyroharmonics of $f$ should be zero at $O(\epsilon^{N-1})$ is
\begin{align}
|B|\text{i}_{\xi}(\tilde{\alpha}_1+\mathbf{d}S)=\tilde{h}_1,
\end{align}
which is precisely Eq.\,(\ref{S}).

\section{\label{app_mu}$\mu_2$ for general magnetic geometry}
The expression for $\mu_2$ we have derived using \emph{VEST} disagrees with the result given in Ref. \onlinecite{WB}. However, upon close examination of the expression for $\mu_2$ given in Ref. \onlinecite{WB}, which is expressed as a sum $\mu_2=\frac{1}{|B|^3}\sum_{n,m=0}^{4}v_\parallel^nv_\perp^ma_{nm}$, we have identified a probable typographical error in the expression for $a_{13}$. Partially and temporarily adopting the notation used in Ref. \onlinecite{WB}, the terms $-\frac{2}{3}(n_1\cdot\nabla\times b)(n_1\cdot\nabla\times n_1-n_2\cdot\nabla\times n_2)$ and $\frac{1}{12}(n_1\cdot\nabla\times b)(n_1\cdot\nabla\times n_1-n_2\cdot\nabla\times n_2)$ were not combined in the obvious way. Given the simplicity of this simplification, it seems likely that one or both of these terms was copied incorrectly. 

Define the vectors $\eta=(\nabla b)\cdot v$ and $\lambda=v\cdot\nabla b$. Also define the scalars $\mu=\frac{(v\times b)\cdot(v\times b)}{2|B|}$, $v_\parallel=b\cdot v$, and $\gamma=v\cdot\nabla\ln|B|$. The expression for $\mu_2$ we have derived and verified using \emph{VEST} is 
\begin{align}
&\mu_2=-\frac{71\mu}{128|B|^2}\eta\cdot\eta-\frac{5v_\parallel\mu}{3|B|^2}\lambda\cdot\nabla\ln|B|\\
&+\frac{10 v_\parallel\mu}{3|B|^2}\nabla\ln|B|\cdot\eta-\frac{715v_\parallel\mu}{192|B|^2}\kappa\cdot\eta+\frac{71\mu}{128|B|^2}(\kappa\cdot v)^2\nonumber\\
&-\frac{3\mu}{2|B|^2}\gamma^2+\frac{71v_\parallel\mu}{64|B|^2}\lambda\cdot\kappa-\frac{7\mu}{128|B|^2}\lambda\cdot\lambda\nonumber\\
&+\frac{25\mu}{64|B|^2}\lambda\cdot\eta+\frac{5v_\parallel}{12|B|^3}(\kappa\cdot v)(\lambda\cdot v)-\frac{2v_\parallel}{3|B|^3}(\gamma)(\lambda\cdot v)\nonumber\\
&+\frac{1}{8|B|^3}(\lambda\cdot v)^2+\frac{217v_\parallel\mu}{32|B|^2}(\kappa\cdot v)(\nabla\cdot b)-\frac{5v_\parallel\mu}{3|B|^2}(\gamma)(\nabla\cdot b)\nonumber\\
&+\frac{23\mu}{32|B|^2}(\lambda\cdot v)(\nabla\cdot b)-\frac{5v_\parallel\mu}{6|B|^2}\nabla^2(b\cdot v)+\frac{5v_\parallel\mu}{6|B|^2}v\cdot\nabla(\nabla\cdot b)\nonumber\\
&+\frac{\mu}{2|B|^3}vv:\nabla\nabla|B|+\frac{5v_\parallel\mu}{6|B|^2}bb:\nabla\nabla(b\cdot v)\nonumber\\
&+\frac{v_\parallel}{6|B|^3}vv:\nabla\nabla(b\cdot v)-\frac{25v_\parallel^2\mu}{12|B|^2}b\cdot\nabla(\nabla\cdot b)-\frac{5v_\parallel^2}{8|B|^3}\eta\cdot\eta\nonumber\\
&+\frac{20v_\parallel^2\mu}{3|B|^2}\kappa\cdot\nabla\ln|B|-\frac{3013v_\parallel^2\mu}{384|B|^2}\kappa\cdot\kappa+\frac{5v_\parallel^2}{6|B|^3}(\kappa\cdot v)^2\nonumber\\
&-\frac{5 v_\parallel^2}{6|B|^3}(\kappa\cdot v)(\gamma)-\frac{29v_\parallel^2}{24|B|^3}\lambda\cdot\lambda+\frac{5v_\parallel^2}{2|B|^3}\lambda\cdot\eta\nonumber\\
&-\frac{5v_\parallel^2}{12|B|^3}(\lambda\cdot v)(\nabla\cdot b)-\frac{25v_\parallel^2\mu}{24|B|^2}(\nabla\cdot b)^2+\frac{55v_\parallel^2\mu}{24|B|^2}\text{Tr}(\nabla b\cdot(\nabla b)^T)\nonumber\\
&-\frac{35v_\parallel^2\mu}{8|B|^2}\text{Tr}(\nabla b\cdot\nabla b)+\frac{5v_\parallel^2}{12|B|^3}bv:\nabla\nabla(b\cdot v)+\frac{5v_\parallel^3}{3|B|^3}\lambda\cdot\kappa\nonumber\\
&+\frac{5v_\parallel^3}{12|B|^3}(\kappa\cdot v)(\nabla\cdot b)+\frac{5v_\parallel^3}{12|B|^3}bb:\nabla\nabla(b\cdot v)+\frac{25v_\parallel^4}{24|B|^3}\kappa\cdot\kappa\nonumber\\
&-\frac{5\mu^2}{4|B|}b\cdot\nabla(\nabla\cdot b)-\frac{5\mu^2}{4|B|^2}\nabla^2|B|+\frac{15\mu^2}{4|B|}\nabla\ln|B|\cdot\nabla\ln|B|\nonumber\\
&-\frac{5\mu^2}{4|B|}\kappa\cdot\nabla\ln|B|-\frac{11\mu^2}{64|B|}\kappa\cdot\kappa-\frac{133\mu^2}{32|B|}(\nabla\cdot b)^2\nonumber\\
&+\frac{11\mu^2}{64|B|}\text{Tr}(\nabla b\cdot(\nabla b)^T)-\frac{5\mu^2}{64|B|}\text{Tr}(\nabla b\cdot\nabla b).\nonumber
\end{align}
In this expression, $x$ and $v$ are the untransformed position and velocity variables. In particular, in the coordinate system used in this expression, the Lorentz force takes the usual form $\dot{v}=|B|v\times b$ and $\dot{x}=\epsilon v$.
%\bibliography{sample_bib_file}

\begin{thebibliography}{29}%
\makeatletter
\providecommand \@ifxundefined [1]{%
 \@ifx{#1\undefined}
}%
\providecommand \@ifnum [1]{%
 \ifnum #1\expandafter \@firstoftwo
 \else \expandafter \@secondoftwo
 \fi
}%
\providecommand \@ifx [1]{%
 \ifx #1\expandafter \@firstoftwo
 \else \expandafter \@secondoftwo
 \fi
}%
\providecommand \natexlab [1]{#1}%
\providecommand \enquote  [1]{``#1''}%
\providecommand \bibnamefont  [1]{#1}%
\providecommand \bibfnamefont [1]{#1}%
\providecommand \citenamefont [1]{#1}%
\providecommand \href@noop [0]{\@secondoftwo}%
\providecommand \href [0]{\begingroup \@sanitize@url \@href}%
\providecommand \@href[1]{\@@startlink{#1}\@@href}%
\providecommand \@@href[1]{\endgroup#1\@@endlink}%
\providecommand \@sanitize@url [0]{\catcode `\\12\catcode `\$12\catcode
  `\&12\catcode `\#12\catcode `\^12\catcode `\_12\catcode `\%12\relax}%
\providecommand \@@startlink[1]{}%
\providecommand \@@endlink[0]{}%
\providecommand \url  [0]{\begingroup\@sanitize@url \@url }%
\providecommand \@url [1]{\endgroup\@href {#1}{\urlprefix }}%
\providecommand \urlprefix  [0]{URL }%
\providecommand \Eprint [0]{\href }%
\providecommand \doibase [0]{http://dx.doi.org/}%
\providecommand \selectlanguage [0]{\@gobble}%
\providecommand \bibinfo  [0]{\@secondoftwo}%
\providecommand \bibfield  [0]{\@secondoftwo}%
\providecommand \translation [1]{[#1]}%
\providecommand \BibitemOpen [0]{}%
\providecommand \bibitemStop [0]{}%
\providecommand \bibitemNoStop [0]{.\EOS\space}%
\providecommand \EOS [0]{\spacefactor3000\relax}%
\providecommand \BibitemShut  [1]{\csname bibitem#1\endcsname}%
\let\auto@bib@innerbib\@empty
%</preamble>
\bibitem [{\citenamefont {Jenko}\ \emph {et~al.}(2000)\citenamefont {Jenko},
  \citenamefont {Dorland}, \citenamefont {Kotschenreuther},\ and\ \citenamefont
  {Rogers}}]{jenko}%
  \BibitemOpen
  \bibfield  {author} {\bibinfo {author} {\bibfnamefont {F.}~\bibnamefont
  {Jenko}}, \bibinfo {author} {\bibfnamefont {W.}~\bibnamefont {Dorland}},
  \bibinfo {author} {\bibfnamefont {M.}~\bibnamefont {Kotschenreuther}}, \ and\
  \bibinfo {author} {\bibfnamefont {B.~N.}\ \bibnamefont {Rogers}},\
  }\href@noop {} {\bibfield  {journal} {\bibinfo  {journal} {Phys. Plasmas}\
  }\textbf {\bibinfo {volume} {7}},\ \bibinfo {pages} {1904} (\bibinfo {year}
  {2000})}\BibitemShut {NoStop}%
\bibitem [{\citenamefont {Dif-Pradalier}\ \emph {et~al.}(2011)\citenamefont
  {Dif-Pradalier}, \citenamefont {Diamond}, \citenamefont {Grandgirard},
  \citenamefont {Sarazin}, \citenamefont {Abiteboul}, \citenamefont {Garbet},
  \citenamefont {Ghendrih}, \citenamefont {Latu}, \citenamefont {Strugarek},
  \citenamefont {Ku}, ,\ and\ \citenamefont {Chang}}]{xgc1}%
  \BibitemOpen
  \bibfield  {author} {\bibinfo {author} {\bibfnamefont {G.}~\bibnamefont
  {Dif-Pradalier}}, \bibinfo {author} {\bibfnamefont {P.~H.}\ \bibnamefont
  {Diamond}}, \bibinfo {author} {\bibfnamefont {V.}~\bibnamefont
  {Grandgirard}}, \bibinfo {author} {\bibfnamefont {Y.}~\bibnamefont
  {Sarazin}}, \bibinfo {author} {\bibfnamefont {J.}~\bibnamefont {Abiteboul}},
  \bibinfo {author} {\bibfnamefont {X.}~\bibnamefont {Garbet}}, \bibinfo
  {author} {\bibfnamefont {P.}~\bibnamefont {Ghendrih}}, \bibinfo {author}
  {\bibfnamefont {G.}~\bibnamefont {Latu}}, \bibinfo {author} {\bibfnamefont
  {A.}~\bibnamefont {Strugarek}}, \bibinfo {author} {\bibfnamefont
  {S.}~\bibnamefont {Ku}}, , \ and\ \bibinfo {author} {\bibfnamefont {C.~S.}\
  \bibnamefont {Chang}},\ }\href@noop {} {\bibfield  {journal} {\bibinfo
  {journal} {Phys. Plasmas}\ }\textbf {\bibinfo {volume} {18}},\ \bibinfo
  {pages} {062309} (\bibinfo {year} {2011})}\BibitemShut {NoStop}%
\bibitem [{\citenamefont {Northrop}(1963)}]{northrop}%
  \BibitemOpen
  \bibfield  {author} {\bibinfo {author} {\bibfnamefont {T.~G.}\ \bibnamefont
  {Northrop}},\ }\href@noop {} {\emph {\bibinfo {title} {The Adiabatic Motion
  of Charged Particles}}},\ Interscience tracts on physics and astronomy\
  (\bibinfo  {publisher} {Interscience Publishers},\ \bibinfo {year}
  {1963})\BibitemShut {NoStop}%
\bibitem [{\citenamefont {Brizard}\ and\ \citenamefont {Tronko}(2012)}]{BrTr}%
  \BibitemOpen
  \bibfield  {author} {\bibinfo {author} {\bibfnamefont {A.~J.}\ \bibnamefont
  {Brizard}}\ and\ \bibinfo {author} {\bibfnamefont {N.}~\bibnamefont
  {Tronko}},\ }\href@noop {} {} (\bibinfo {year} {2012}),\ \Eprint
  {http://arxiv.org/abs/arXiv:1205.5772} {arXiv:1205.5772} \BibitemShut
  {NoStop}%
\bibitem [{\citenamefont {Littlejohn}(1984)}]{GGI}%
  \BibitemOpen
  \bibfield  {author} {\bibinfo {author} {\bibfnamefont {R.~G.}\ \bibnamefont
  {Littlejohn}},\ }in\ \href@noop {} {\emph {\bibinfo {booktitle} {Fluids and
  Plasmas: Geometry and Dynamics}}},\ \bibinfo {series} {Contemporary
  mathematics}, Vol.~\bibinfo {volume} {28},\ \bibinfo {editor} {edited by\
  \bibinfo {editor} {\bibfnamefont {J.~E.}\ \bibnamefont {Marsden}}}\ (\bibinfo
   {publisher} {American Mathematical Society},\ \bibinfo {year} {1984})\ pp.\
  \bibinfo {pages} {151--167}\BibitemShut {NoStop}%
\bibitem [{\citenamefont {Brizard}(1990)}]{Brizard_thesis}%
  \BibitemOpen
  \bibfield  {author} {\bibinfo {author} {\bibfnamefont {A.}~\bibnamefont
  {Brizard}},\ }\emph {\bibinfo {title} {Nonlinear Gyrokinetic Tokamak
  Physics}},\ \href@noop {} {\bibinfo {type} {{PhD} dissertation}},\ \bibinfo
  {school} {Princeton University}, \bibinfo {address} {Department of
  Astrophysical Sciences} (\bibinfo {year} {1990})\BibitemShut {NoStop}%
\bibitem [{\citenamefont {Krommes}(2012)}]{krommes12}%
  \BibitemOpen
  \bibfield  {author} {\bibinfo {author} {\bibfnamefont {J.}~\bibnamefont
  {Krommes}},\ }\href@noop {} {\bibfield  {journal} {\bibinfo  {journal} {Ann.
  Rev. Fluid Mech.}\ }\textbf {\bibinfo {volume} {44}},\ \bibinfo {pages} {175}
  (\bibinfo {year} {2012})}\BibitemShut {NoStop}%
\bibitem [{\citenamefont {Parra}\ and\ \citenamefont {Catto}(2010)}]{PaCa10}%
  \BibitemOpen
  \bibfield  {author} {\bibinfo {author} {\bibfnamefont {F.~I.}\ \bibnamefont
  {Parra}}\ and\ \bibinfo {author} {\bibfnamefont {P.~J.}\ \bibnamefont
  {Catto}},\ }\href@noop {} {\bibfield  {journal} {\bibinfo  {journal} {Phys.
  Plasmas}\ }\textbf {\bibinfo {volume} {17}},\ \bibinfo {pages} {056106}
  (\bibinfo {year} {2010})}\BibitemShut {NoStop}%
\bibitem [{\citenamefont {Littlejohn}(1981)}]{littlejohn81}%
  \BibitemOpen
  \bibfield  {author} {\bibinfo {author} {\bibfnamefont {R.~G.}\ \bibnamefont
  {Littlejohn}},\ }\href@noop {} {\bibfield  {journal} {\bibinfo  {journal}
  {Phys.\ Fluids}\ }\textbf {\bibinfo {volume} {24}},\ \bibinfo {pages} {1730}
  (\bibinfo {year} {1981})}\BibitemShut {NoStop}%
\bibitem [{\citenamefont {Littlejohn}(1983)}]{littlejohn83}%
  \BibitemOpen
  \bibfield  {author} {\bibinfo {author} {\bibfnamefont {R.~G.}\ \bibnamefont
  {Littlejohn}},\ }\href@noop {} {\bibfield  {journal} {\bibinfo  {journal} {J.
  Plasma Phys.}\ }\textbf {\bibinfo {volume} {29}},\ \bibinfo {pages} {111}
  (\bibinfo {year} {1983})}\BibitemShut {NoStop}%
\bibitem [{\citenamefont {Kruskal}(1962)}]{kruskal62}%
  \BibitemOpen
  \bibfield  {author} {\bibinfo {author} {\bibfnamefont {M.}~\bibnamefont
  {Kruskal}},\ }\href@noop {} {\bibfield  {journal} {\bibinfo  {journal} {J.
  Math. Phys.}\ }\textbf {\bibinfo {volume} {3}},\ \bibinfo {pages} {806}
  (\bibinfo {year} {1962})}\BibitemShut {NoStop}%
\bibitem [{\citenamefont {Omohundro}(1986)}]{omohundro}%
  \BibitemOpen
  \bibfield  {author} {\bibinfo {author} {\bibfnamefont {S.~M.}\ \bibnamefont
  {Omohundro}},\ }\href@noop {} {\emph {\bibinfo {title} {Geometric
  Perturbation Theory in Physics}}}\ (\bibinfo  {publisher} {World Scientific
  Publishing Co. Pte. Ltd., Singapore},\ \bibinfo {year} {1986})\BibitemShut
  {NoStop}%
\bibitem [{\citenamefont {Squire}, \citenamefont {Burby},\ and\ \citenamefont
  {Qin}(2013)}]{SqBuQi13}%
  \BibitemOpen
  \bibfield  {author} {\bibinfo {author} {\bibfnamefont {J.}~\bibnamefont
  {Squire}}, \bibinfo {author} {\bibfnamefont {J.~W.}\ \bibnamefont {Burby}}, \
  and\ \bibinfo {author} {\bibfnamefont {H.}~\bibnamefont {Qin}},\ }\href@noop
  {} {\enquote {\bibinfo {title} {Vest: abstract vector calculus simplification
  in mathematica},}\ } (\bibinfo {year} {2013}),\ \bibinfo {note} {{Comp. Phys.
  Comm.} (submitted)}\BibitemShut {NoStop}%
\bibitem [{\citenamefont {Kruskal}(1958)}]{kruskal58}%
  \BibitemOpen
  \bibfield  {author} {\bibinfo {author} {\bibfnamefont {M.}~\bibnamefont
  {Kruskal}},\ }\href@noop {} {\enquote {\bibinfo {title} {The gyration of a
  charged particle},}\ }\bibinfo {type} {Project Matterhorn Report}\ \bibinfo
  {number} {PM-S-33 (NYO-7903)}\ (\bibinfo  {institution} {Princeton
  University},\ \bibinfo {year} {1958})\BibitemShut {NoStop}%
\bibitem [{\citenamefont {Larsson}(1986)}]{larsson}%
  \BibitemOpen
  \bibfield  {author} {\bibinfo {author} {\bibfnamefont {J.}~\bibnamefont
  {Larsson}},\ }\href@noop {} {\bibfield  {journal} {\bibinfo  {journal}
  {Physica Scripta}\ }\textbf {\bibinfo {volume} {33}},\ \bibinfo {pages} {342}
  (\bibinfo {year} {1986})}\BibitemShut {NoStop}%
\bibitem [{\citenamefont {Abraham}\ and\ \citenamefont {Marsden}(1978)}]{FoM}%
  \BibitemOpen
  \bibfield  {author} {\bibinfo {author} {\bibfnamefont {R.}~\bibnamefont
  {Abraham}}\ and\ \bibinfo {author} {\bibfnamefont {J.}~\bibnamefont
  {Marsden}},\ }\href@noop {} {\emph {\bibinfo {title} {Foundations of
  Mechanics}}},\ AMS Chelsea publishing\ (\bibinfo  {publisher} {AMS Chelsea
  Pub./American Mathematical Society},\ \bibinfo {year} {1978})\BibitemShut
  {NoStop}%
\bibitem [{\citenamefont {Lee}(2003)}]{smooth_manifolds}%
  \BibitemOpen
  \bibfield  {author} {\bibinfo {author} {\bibfnamefont {J.~M.}\ \bibnamefont
  {Lee}},\ }\href@noop {} {\emph {\bibinfo {title} {Introduction to Smooth
  Manifolds}}},\ Graduate Texts in Mathematics\ (\bibinfo  {publisher}
  {Springer-Verlag New York},\ \bibinfo {year} {2003})\BibitemShut {NoStop}%
\bibitem [{\citenamefont {Burby}\ and\ \citenamefont {Qin}(2012)}]{BuQi12}%
  \BibitemOpen
  \bibfield  {author} {\bibinfo {author} {\bibfnamefont {J.~W.}\ \bibnamefont
  {Burby}}\ and\ \bibinfo {author} {\bibfnamefont {H.}~\bibnamefont {Qin}},\
  }\href@noop {} {\bibfield  {journal} {\bibinfo  {journal} {Phys. Plasmas}\
  }\textbf {\bibinfo {volume} {19}},\ \bibinfo {pages} {052106} (\bibinfo
  {year} {2012})}\BibitemShut {NoStop}%
\bibitem [{\citenamefont {Qin}(2005)}]{qinRINGS}%
  \BibitemOpen
  \bibfield  {author} {\bibinfo {author} {\bibfnamefont {H.}~\bibnamefont
  {Qin}},\ }in\ \href@noop {} {\emph {\bibinfo {booktitle} {Fields Institute
  Communications}}},\ Vol.~\bibinfo {volume} {46},\ \bibinfo {editor} {edited
  by\ \bibinfo {editor} {\bibfnamefont {D.}~\bibnamefont {Levermore}}, \bibinfo
  {editor} {\bibfnamefont {T.}~\bibnamefont {Passot}}, \bibinfo {editor}
  {\bibfnamefont {C.}~\bibnamefont {Sulem}}, \ and\ \bibinfo {editor}
  {\bibfnamefont {P.}~\bibnamefont {Sulem}}}\ (\bibinfo  {publisher} {American
  Mathematical Society},\ \bibinfo {year} {2005})\ pp.\ \bibinfo {pages}
  {171--192}\BibitemShut {NoStop}%
\bibitem [{Note1()}]{Note1}%
  \BibitemOpen
  \bibinfo {note} {For some quick intuition regarding this ansatz, recall that
  the time-advance map associated to an arbitrary vector field $Y:M\rightarrow
  TM$ is given by $\protect \qopname \relax o{exp}(t Y)$. Thus, given $x\in M$,
  $T_\epsilon (x)$ is found by sequentially flowing along the vector fields
  $G_n(\epsilon )$ for $-1$ unit of time each, starting from $x$.}\BibitemShut
  {Stop}%
\bibitem [{\citenamefont {L.~De~Guillebon}(2013)}]{guillebon}%
  \BibitemOpen
  \bibfield  {author} {\bibinfo {author} {\bibfnamefont {L. De}\ \bibnamefont
  {Guillebon}}\ and\ \bibinfo {author} {\bibfnamefont {M.}~\bibnamefont {Vittot}},\ }\href@noop {} {} (\bibinfo {year} {2013}),\ \Eprint
  {http://arxiv.org/abs/arXiv:1211.5792} {arXiv:1211.5792} \BibitemShut
  {NoStop}%
\bibitem [{\citenamefont {Bogoliubov}(1961)}]{KBM}%
  \BibitemOpen
  \bibfield  {author} {\bibinfo {author} {\bibfnamefont {N.}~\bibnamefont
  {Bogoliubov}},\ }\href@noop {} {\emph {\bibinfo {title} {Asymptotic Methods
  in the Theory of Non-Linear Oscillations}}},\ International monographs on
  advanced mathematics and physics\ (\bibinfo  {publisher} {Hindustan},\
  \bibinfo {year} {1961})\BibitemShut {NoStop}%
\bibitem [{\citenamefont {Littlejohn}(1982)}]{littlejohn82}%
  \BibitemOpen
  \bibfield  {author} {\bibinfo {author} {\bibfnamefont {R.~G.}\ \bibnamefont
  {Littlejohn}},\ }\href@noop {} {\bibfield  {journal} {\bibinfo  {journal} {J.
  Math. Phys.}\ }\textbf {\bibinfo {volume} {23}},\ \bibinfo {pages} {742}
  (\bibinfo {year} {1982})}\BibitemShut {NoStop}%
\bibitem [{Note2()}]{Note2}%
  \BibitemOpen
  \bibinfo {note} {Demanding the transformed $\protect \mathbf {d}\vartheta
  _\epsilon $ to be gyrosymmetric is equivalent to demanding that the
  transformed $\vartheta _\epsilon $ be equal to the sum of a gyrosymmetric
  one-form and an arbitrary closed one-form. This follows from the fact that
  $\protect \mathbf {d}\vartheta _\epsilon ^\prime $ is left unchanged upon
  replacing $\vartheta _\epsilon ^\prime $ with $\vartheta _\epsilon ^\prime
  +\alpha $ for an arbitrary closed one-form $\alpha $, i.e. $\protect \mathbf
  {d}\alpha =0$}\BibitemShut {NoStop}%
\bibitem [{Note3()}]{Note3}%
  \BibitemOpen
  \bibinfo {note} {F stands for ``fibered''.}\BibitemShut {Stop}%
\bibitem [{Note4()}]{Note4}%
  \BibitemOpen
  \bibinfo {note} {In spite of its simplicity and utility, this formula only
  seems to have been noticed recently. It can be found in the literature in
  Ref. \protect \rev@citealpnum {BuQi12}. It was also independently discovered
  by Zhi Yu\cite {zhi}, but never published.}\BibitemShut {Stop}%
\bibitem [{Note5()}]{Note5}%
  \BibitemOpen
  \bibinfo {note} {ND stands for ``non-degenerate''.}\BibitemShut {Stop}%
\bibitem [{\citenamefont {Weyssow}\ and\ \citenamefont {Balescu}(1986)}]{WB}%
  \BibitemOpen
  \bibfield  {author} {\bibinfo {author} {\bibfnamefont {B.}~\bibnamefont
  {Weyssow}}\ and\ \bibinfo {author} {\bibfnamefont {R.}~\bibnamefont
  {Balescu}},\ }\href@noop {} {\bibfield  {journal} {\bibinfo  {journal} {J.
  Plasma Phys.}\ }\textbf {\bibinfo {volume} {35}},\ \bibinfo {pages} {449}
  (\bibinfo {year} {1986})}\BibitemShut {NoStop}%
\bibitem [{\citenamefont {Parra}\ and\ \citenamefont {Calvo}(2011)}]{PnC}%
  \BibitemOpen
  \bibfield  {author} {\bibinfo {author} {\bibfnamefont {F.}~\bibnamefont
  {Parra}}\ and\ \bibinfo {author} {\bibfnamefont {I.}~\bibnamefont
  {Calvo}},\ }\href@noop {} {\bibfield  {journal} {\bibinfo  {journal} {Plasma Phys.
  Control. Fusion}\ }\textbf {\bibinfo {volume} {53}},\ \bibinfo {pages} {045001}
  (\bibinfo {year} {2011})}\BibitemShut {NoStop}%
\bibitem [{\citenamefont {Yu}(2011)}]{zhi}%
  \BibitemOpen
  \bibfield  {author} {\bibinfo {author} {\bibfnamefont {Z.}~\bibnamefont
  {Yu}},\ }\href@noop {} {}\bibinfo {howpublished} {private communication}
  (\bibinfo {year} {2011}),\ \bibinfo {note}
  {http://meetings.aps.org/link/BAPS.2011.DPP.TO4.15}\BibitemShut {NoStop}%
\end{thebibliography}

%%%%%%%%%%%%%%%%%%%%%%%%%%%%%%%%%%%%%

%% put content of bib file here when ready to submit
\providecommand{\noopsort}[1]{}\providecommand{\singleletter}[1]{#1}%
%

%%%%%%%%%%%%%%%%%%%%%%%%%%%%%%%%%%%%

\end{document}